\documentclass[%
 reprint,
 amsmath,amssymb,
 aps,
pra,
floatfix,
]{revtex4-1}

\usepackage{ifpdf}
\usepackage{amsmath} 
\usepackage{amssymb}
\usepackage{amsfonts}
\usepackage{braket}
\usepackage{color}
\usepackage[colorlinks=true,linkcolor=blue,citecolor=blue,breaklinks]{hyperref}
\usepackage{graphicx}
\usepackage{dcolumn}
\usepackage{bm}
\usepackage{array}

\newcommand{\hc}{\text{H.c.}}
\newcommand{\la}{\langle}
\newcommand{\ra}{\rangle}

\newcommand{\re}{\text{Re}}
\newcommand{\I}{\text{I}}
\newcommand{\II}{\text{II}}
\newcommand{\III}{\text{III}}
\newcommand{\pathD}{\mathfrak{D}}
\newcommand{\ii}{\mathrm{i}\,}

\usepackage{todonotes}

\begin{document}

\title{Out-of-time-ordered correlators in short-range and long-range hard-core boson models and in the Luttinger liquid model}

\author{Cheng-Ju Lin}
\affiliation{Department of Physics and Institute for Quantum Information and Matter, California Institute of Technology, Pasadena, CA 91125, USA}
\author{Olexei I. Motrunich}
\affiliation{Department of Physics and Institute for Quantum Information and Matter, California Institute of Technology, Pasadena, CA 91125, USA}


\date{\today}

\begin{abstract}
We study out-of-time-ordered correlators (OTOCs) in hard-core boson models with short-range and long-range hopping and compare the results to the OTOCs in the Luttinger-liquid model.
For density-density correlations, a related expectation value of the squared commutator starts at zero and decays back to zero after the passage of the wavefront in all three models, while the wavefront broadens as $t^{1/3}$ in the short-range model and shows no broadening in the long-range model and the Luttinger-liquid model. 
For the boson creation operator, the corresponding commutator function shows saturation inside the light cone in all three models, with similar wavefront behavior as in the density-density commutator function, despite the presence of a nonlocal string in terms of Jordan-Wigner fermions.
For the long-range model and the Luttinger-liquid model, the commutator function decays as a power law outside the light cone in the long time regime when following different fixed-velocity rays.
In all cases, the OTOCs approach their long-time values in a power-law fashion, with different exponents for different observables and short-range versus long-range cases.
Our long-range model appears to capture exponents in the Luttinger liquid model (which are found to be independent of the Luttinger parameter in the model).
This conclusion also comes to bear on the OTOC calculations in conformal field theories, which we propose correspond to long-ranged models.

\end{abstract}

\maketitle


\section{\label{sec:intro} Introduction}
Recently, out-of-time-ordered correlators (OTOCs) have emerged as providing a diagnostic for quantum chaos and information scrambling\cite{a.i.larkinQuasiclassical1969,shenkerBlack2014,alexeikitaevSimple2015,robertsLocalized2015,robertsDiagnosing2015,kitaevSoft2018,maldacenaBound2016,guFractional2016,hosurChaos2016,robertsLiebRobinson2016,aleinerMicroscopic2016,chowdhuryOnset2017,patelQuantum2017,patelQuantum2017a,wermanQuantum,cotlerChaos2017a,chapmanClassical2018,yungerhalpernJarzynskilike2017,yungerhalpernQuasiprobability2018,rozenbaumLyapunov2017,swingleSlow2017,luitzInformation2017a,shenOutoftimeorder2017,fanOutoftimeorder2017,slagleOutoftimeorder2017,huangOutoftimeordered2017,chenOutoftimeorder2017,kukuljanWeak2017,hashimotoOutoftimeorder2017,cotlerOutoftimeorder,doraOutofTimeOrdered2017,bohrdtScrambling2017,scaffidiSemiclassical,linOutoftimeordered2018,xuAccessing,xuLocality,alonsoOutoftimeorderedcorrelator,jianUniversal,knapEntanglement}.
Intense efforts have also been devoted to devise experimental measurements for such a  quantity\cite{yaoInterferometric,garttnerMeasuring2017,liMeasuring2017,swingleResilience2018,dresselStrengthening2018}.
Consider the following (squared) ``commutator function:''
\begin{equation}
C_{WV}(\ell, t) = \frac{1}{2} \left\la [W_0(t), V_\ell]^\dagger [W_0(t), V_\ell] \right\ra ~,
\label{eqn:CWV}
\end{equation}
for an operator $W_0$ at the origin (site $0$) and an operator $V_\ell$ at site $\ell$, where the average $\la A \ra \equiv \text{Tr}[\rho A]/\text{Tr}[\rho]$ is with respect to the Gibbs ensemble $\rho \equiv e^{-\beta H}$.
If we expand the commutator and further assume that $W$ and $V$ are Hermitian and unitary (as is the case when $W$ and $V$ are Pauli matrices in a spin-1/2 system), we can write 
\begin{align}
& C_{WV}(\ell, t) = 1-\re F_{WV}(\ell,t) ~, \\
& F_{WV}(\ell, t) = \la W_0(t) V_\ell W_0(t) V_\ell \ra ~,
\label{eqn:FWV}
\end{align}
where $F_{WV}(\ell, t)$ is called the OTOC due to the unusual time ordering.

One interpretation of the commutator function $C_{WV}(\ell, t)$ is as a quantification of operator spreading\cite{vonkeyserlingkOperator2018,nahumOperator2018,khemaniOperator2018,khemaniVelocitydependent,rakovszkyDiffusive2018,gopalakrishnanOperator2018a}.
Consider the Heisenberg evolution of $W_0(t) = \sum_S a_S(t) S$, where $S$ runs over all Pauli-string operators (e.~g., $\dots \sigma_j^z \sigma_{j\!+\!1}^x \dots$) for spin-$\frac{1}{2}$ systems.
Then $C_{WV}(\ell, t)$ has contributions from Pauli strings that do not commute with $V_\ell$, therefore providing a measurement of the ``shape'' of the operator and its spreading with time.
Strictly speaking, this picture is only valid at infinite temperature.
However, we can still understand $C_{WV}(\ell, t)$ at a finite temperature as measuring the operator spreading averaged over energy eigenstates in the corresponding energy window of the many-body spectrum.

Despite their recent prominence, the OTOCs are difficult to evaluate in quantum many-body systems, and there are many numerical studies on small systems with conflicting interpretations but few rigorous results.
Early analytical calculations were done in conformal field theories~\cite{robertsDiagnosing2015} and Luttinger liquids~\cite{doraOutofTimeOrdered2017}; however, these have Lorentz symmetry, and it is not clear how to connect them to the lattice models.
Our previous study of the quantum Ising chain\cite{linOutoftimeordered2018} provided a non-trivial lattice calculation with results that differed from the Ising conformal field theory (CFT) predictions, particularly near the wavefront and in the long-time approach.
The non-trivial aspect here is that the magnetization observable $\sigma^z$ becomes non-local in terms of the Jordan-Wigner fermions and is known to have different dynamical and thermalization properties from the transverse field observable $\sigma^x$ that is local in the Jordan-Wigner (JW) fermions; in particular, we found that the $\sigma^z$-$\sigma^z$ OTOC has unusual slow $t^{-1/4}$ power-law decay, for which we do not have an analytical understanding.
To shine more light on these issues, in this paper we therefore consider OTOCs in exactly tractable short-range and long-range hardcore boson models, and we compare with the Luttinger liquid model, which corresponds to one of the simplest CFTs.

To set the stage further, we quickly review some current and still developing understanding of the OTOCs.
For systems whose evolution is described by a local Hamiltonian dynamics or a local quantum circuit, a description has recently emerged that operators spread with a front ballistically, with a velocity $v_B$ dubbed ``butterfly velocity.''
For systems governed by a local Hamiltonian, outside the light cone, at very short time, the commutator function exhibits a position-dependent power-law growth in time, which can be understood using Baker-Campbell-Hausdorff expansion of the Heisenberg evolution of operators\cite{robertsLiebRobinson2016,doraOutofTimeOrdered2017,linOutoftimeordered2018,xuAccessing}.
On the other hand, inside the light cone, the saturation of $C(\ell, t)$ [equivalently $F(\ell, t)$ approaching zero] at long time is commonly used as a diagnostic for scrambling or quantum chaos. 

While operators spread with a ballistic velocity, the front itself can broaden.
It has been proposed recently\cite{xuAccessing} that the functional form of the wavefront has a universal description
\begin{equation}\label{eqn:Xu_proposal}
C(\ell, t) \sim \exp\left[ -c \frac{(\ell - v_B t)^{1+p}}{t^p} \right] ~.
\end{equation}
(One has to also carefully specify the window around the wavefront where such a description is valid.)
Another characterization is to examine long-time behavior along fixed-velocity rays\cite{khemaniVelocitydependent}.
For systems governed by local Hamiltonians, outside the light cone, $v > v_B$, one expects
\begin{equation}\label{eqn:Khemani_proposal}
C(\ell = vt, t) \sim \exp[-\lambda(v) t] ~,
\end{equation}
where $\lambda(v)$ is dubbed a ``velocity-dependent'' Lyapunov exponent.

Note that strictly speaking, the above two proposals, Eqs.~(\ref{eqn:Xu_proposal}) and (\ref{eqn:Khemani_proposal}), are describing different asymptotic regimes.
However, if the two descriptions can be connected smoothly, then one obtains $\lambda(v) = c (v - v_B)^{1+p}$.
The exponent $p$ describes wavefront broadening as $\sim t^{p/(1+p)}$\cite{khemaniVelocitydependent}.
For example, for the random circuit model\cite{vonkeyserlingkOperator2018, nahumOperator2018}, we have $p = 1$ corresponding to $\sim t^{1/2}$ spreading.
For models with a noninteracting fermionic quasiparticle description\cite{linOutoftimeordered2018, xuAccessing, khemaniVelocitydependent}, we have $p = 1/2$ and $\sim t^{1/3}$ spreading.
Finally, for one-dimensional (1D) chains of coupled Sachdev-Ye-Kitaev quantum dots and models with a large-$N$ limit, $p = 0$ and the wavefront does not broaden but shows an exponential growth at fixed $\ell$ and increasing $t$, which is reminiscent of the classical chaos---the butterfly effect\cite{maldacenaBound2016,kitaevSoft2018, aleinerMicroscopic2016,Gu2017,xuAccessing, khemaniVelocitydependent}.
While the existence of a well-defined exponential growth regime for local Hamiltonians with bounded local Hilbert spaces is still an outstanding question (with emerging thinking that there is probably no such regime), a recent work\cite{chenMeasuring2018} has reported an exponential growth near the wavefront in spin models with long-range interactions.

It is therefore also interesting to examine how the above descriptions are modified in models with noninteracting fermionic quasiparticles with long-range hopping.
To this end, in this paper, we consider hard-core boson models that have such properties. 
First, we consider hard-core bosons with nearest-neighbor hopping\cite{mccoyStatistical1971,perkTimedependent1977,capelAutocorrelation1977}.
By JW transformation, the model maps to free fermions with nearest-neighbor hopping.
It is known (but not widely appreciated) that at a finite temperature, the dynamics is \emph{not} described by the (linear) Luttinger liquid model\cite{imambekovOnedimensional2012}.
In order to compare with the OTOCs in the Luttinger liquid model, we propose to artificially ``straighten'' the free-fermion dispersion, which leads to our second model:
Such bounded linear dispersion corresponds to long-range hopping of the fermions, or equivalently some specific multi-body interaction of the hard-core bosons.
We then discuss how the commutator functions behave differently compared to the short-range model and the agreements and disagreements between the long-range hopping model and the Luttinger model\cite{haldaneLuttinger1981,haldaneEffective1981,giamarchiQuantum2003}.
In particular, we propose a resolution of the question of which systems are described by the OTOC calculations in the Luttinger model\cite{doraOutofTimeOrdered2017}, which also bears on the OTOC calculations in CFTs.

The paper is organized as follows.
In Sec.~\ref{sec:models}, we define the models we study in this paper.
We then discuss the density-density OTOC in Sec.~\ref{sec:density-density} and boson-boson OTOC in Sec.~\ref{sec:boson_lattice} for the hard-core boson lattice models and in Sec.~\ref{sec:boson_luttinger} for the continuum Luttinger liquid model.
For all cases, we focus on the early-time (well before the wavefront), early-growth (behavior around the wavefront), and long-time (well after the wavefront) behaviors of the OTOCs.
We conclude in Sec.~\ref{sec:conclusion} with some discussion and open questions.
For the readers' convenience, we summarize our results for the three models in the different regimes in Table~\ref{table:summary}.

\begin{table}
Density-density OTOC
\begin{tabular}{| m{2cm} || m{1.5cm} | m{2cm} | m{1.5cm} |}
\hline
  & Early-time & Wavefront Broadening & Long time approach\\
\hline
Model I    & $t^{2\ell}/(\ell!)^2$ & $t^{\frac{1}{3}}$ & $t^{-1}$ \\
Model II   & $t^2/\ell^4$ & No (i.e., $t^0$)  & $t^{-2}$  \\ 
Model III(a) & $t^2/\ell^4$ & No  & $t^{-2}$  \\
Model III(b) & $t^2/\ell^2$ & No  & $t^{-2}$  \\
\hline
\end{tabular}
Boson-boson OTOC
\begin{tabular}{| m{2cm} || m{1.5cm} | m{2cm} | m{1.5cm} |}
\hline
  & Early-time & Wavefront Broadening & Long time approach \\
\hline
Model I    & $t^{2\ell}/(\ell!)^2$ & $t^{\frac{1}{3}}$ & $t^{-\frac{1}{2}}$ \\
Model II   & $t^2/\ell^2$ & No & $t^{-1}$ \\ 
Model III(a) & $t^6/\ell^4$ & No & $t^{-1}$ \\
Model III(b) & $t^2/\ell^2$ & No & $t^{-1}$ \\
\hline
\end{tabular}
\caption{\label{table:summary}
Summary of the main results for the three models in the different regimes.
Model I is the nearest-neighbor lattice model; model II is the long-range hopping lattice model; and model III is the continuum Luttinger liquid model.
The results for model III are quoted with the Luttinger parameter $g=1$ corresponding to noninteracting fermions and (a) cutoff $\Lambda=\pi$ and (b) generic cutoff $\Lambda \neq \pi$.}
\end{table}

\section{Models}\label{sec:models}
In this section, we define more precisely the models we study.
Consider a Hamiltonian defined on lattice sites $i = -L/2 + 1, \dots, L/2$, where we have assumed that the number of sites $L$ is even for simplicity, 
\begin{equation}
H = \sum_{i < j} J_{ij} \left[ b_i^\dagger \left(e^{\ii \pi \sum_{r=i+1}^{j-1} n_r} \right) b_j + \hc \right] - \mu \sum_i n_i ~,
\end{equation}
with open boundary conditions and real couplings $J_{ij}$; also, $n_i \equiv b_i^\dagger b_i$ is the boson number operator.
The boson operators are hard-core bosons commuting on different sites.
The choice of the Hamiltonian is such that under the JW transformation $b_j = (\prod_{r = -L/2+1}^{j-1} e^{\ii \pi n_r}) c_j$, the Hamiltonian becomes
\begin{equation}
H = \sum_{i < j} J_{ij} \left( c_i^\dagger c_j + \hc \right) -\mu \sum_i n_i ~.
\end{equation}

The first model we consider is the ``short-range hopping'' model (various quantities defined and calculated in this model will be labeled by ``$\I$''),
defined by 
\begin{equation}
J_{ij}^\I = -\frac{v_B}{2} (\delta_{i=j\!-\!1} + \delta_{i=j\!+\!1}) ~.
\end{equation}
The Hamiltonian can be diagonalized by the transformation 
$c_{k} = \sqrt{\frac{2}{L+1}} \sum_{j = -L/2+1}^{L/2} \sin(k \bar{j}) c_j$, where $\bar{j}\equiv j\!+\!L/2$ and $\{ k\!=\!\frac{n\pi}{L+1}$, $n = 1, \dots, L \}$.
We obtain $H = \sum_{k} \epsilon^\I(k) c_{k}^\dagger c_{k}$, where the dispersion $\epsilon^\I(k) = -v_B\cos(k) - \mu$.
The coupling is chosen such that the maximum group velocity $v_\text{max} = \text{max} |\partial \epsilon_k/\partial k| = v_B$.
We choose $v_B = 1$ as our energy unit and throughout set $\hbar = 1$.

For the second model (with quantities labeled by ``$\II$,''), we artificially ``straighten'' the dispersion, making it as $\epsilon^\II(k) = J |k| - \mu$ for $k \in [-\pi, \pi]$.
For general $k' \not \in [-\pi, \pi]$, $\epsilon^\II(k') = \epsilon^\II(k)$ where $k = k' + 2\pi m$, with $m$ some integer such that $k \in [-\pi, \pi]$.
In real space, $J_{ij}^\II = \frac{2}{L+1} \sum_{k_n} (\epsilon^\II(k_n) + \mu) \sin(k_n \bar{i}) \sin (k_n \bar{j})$.
In the thermodynamic limit $L \to \infty$, for points in the bulk, we have 
\begin{equation}
J_{ij}^\II = \frac{v_B}{\pi} \frac{[(-1)^{|i-j|} - 1]}{|i-j|^2} ~.
\end{equation}
We will focus on the cases where $\mu$ is tuned such that the ground state is in the gapless phase (quasi-long-range ordered).

Finally, we will also compare the results to the Luttinger liquid model\cite{haldaneLuttinger1981, haldaneEffective1981, giamarchiQuantum2003} (quantities labeled by ``$\III$''), defined as
\begin{equation}
H^\III = \frac{v_B}{2\pi} \int_0^L dx \left[ g (\pi \hat{\Pi})^2 + \frac{1}{g} (\partial_x \hat{\theta})^2 \right] ~,
\end{equation}
 where we set the characteristic velocity as $v_B$. 
As we will see later, this will indeed be the butterfly velocity.
$\hat{\theta}(x)$ is related to the density operator defined as 
\begin{equation}\label{eqn:density_op}
n(x) = d_0 + \rho_0(x) + d_2 W(x) ~,
\end{equation}
where $\rho_0(x) \equiv -\partial_x \hat{\theta}(x)/\pi$ and 
\begin{equation}
W(x) \equiv e^{i 2\pi d_0 x} V_{-2}(x) + e^{-i 2\pi d_0 x} V_2(x) ~,
\end{equation}
and $V_m(x) \equiv e^{i m \hat{\theta}(x)}$ is the vertex operator, while $d_0 = k_F/\pi$ is the density of the system, and $d_2$ is some constant to be determined.
$\hat{\Pi}(x)$ is the conjugate momentum to $\hat{\theta}(x)$, satisfying $[\hat{\Pi}(x), \hat{\theta}(x')] = -\ii \delta(x-x')$.
Since we are studying bosonic models, we will also consider the boson creation field
\begin{equation}
\psi_B^\dagger(x) \sim e^{\ii \hat{\phi}(x)} ~,
\end{equation}
where the field $\hat{\phi}(x)$ is the phase field defined by the relation $\hat{\Pi}(x) = -\partial_x \hat{\phi}/\pi$.
To be concrete, here we use periodic boundary conditions, $\hat{\theta}(x+L) = \hat{\theta}(x)$.
However, as most of our calculations will be taken in the thermodynamic limit, the choice of the boundary conditions will not matter.

The Luttinger liquid model can be diagonalized as follows.
We define Fourier modes 
$\theta_k = \frac{1}{\sqrt{L}} \int_0^L dx e^{-i k x} \hat{\theta}(x)$ and 
$\Pi_k = \frac{1}{\sqrt{L}} \int_0^L dx e^{-i k x} \hat{\Pi}(x)$,
and find 
$H^\III = \frac{v_B}{2\pi} \sum_k (\pi^2 g \Pi_{-k} \Pi_k + \frac{k^2}{g} \theta_{-k} \theta_k)$.
We can identify $\omega_k = v_B |k|$ and $m = \frac{1}{\pi v_B g}$ as in a harmonic oscillator.
We now define ladder operators $b_k = \sqrt{\frac{m \omega_k}{2}} (\theta_k + \frac{\ii}{m \omega_k} \Pi_{k})$ and $b_k^\dagger = \sqrt{\frac{m \omega_k}{2}} (\theta_{-k} - \frac{\ii}{m \omega_k} \Pi_{-k})$, which satisfy canonical boson commutation relations $[b_k, b_{k'}] = 0$, $[b_k, b_{k'}^\dagger] = \delta_{k,k'}$.
The Hamiltonian becomes $H^\III = \sum_k \omega_k (b_k^\dagger b_k + \frac{1}{2})$, and the fields $\hat{\theta}(x)$ and $\hat{\phi}(x)$ can be expressed as linear combinations of the eigenmode operators $b_k$ and $b_k^\dagger$.

\section{Density-density OTOC}
\label{sec:density-density}
We first consider the density-density OTOC $F_{nn}(\ell, t)$ [and the squared commutator $C_{nn}(\ell, t)$], where $n_i \equiv b_i^\dagger b_i$.
The density-density OTOC in fact can be calculated analytically and relatively easily in the lattice models since the operators can be expressed using few JW fermion operators.
Detailed calculations are presented in Appendix~\ref{app:density-density_lattice} for the lattice models and Appendix~\ref{app:density_luttinger} for the Luttinger liquid model.

\subsection{Density-density OTOC in the lattice models}
In the lattice models, we find
\begin{align}
C_{nn}(\ell, t) = |A(\ell, t)|^2 \Big( & [\la n_\ell \ra + \la n_0 \ra]/2 - \la n_\ell \ra \la n_0 \ra \nonumber \\
& -\re\left[ \la c_0^\dagger(t) c_\ell \ra \la c_0(t) c_\ell^\dagger \ra \right] \Big) ~, 
\end{align}
where
\begin{align}
A(\ell, t) \equiv \int_{-\pi}^\pi \frac{dk}{2\pi} e^{\ii (k \ell - \epsilon_k t)} ~
\end{align}
is a specific fermion evolution function [which appears, e.g., in the anti-commutator for the fermion fields, $\{ c_0(t), c_\ell^\dagger \} = A(\ell, t)$].
At infinite temperature ($\beta = 0$), $C_{nn}(\ell, t) = \frac{1}{4} (|A(\ell, t)|^2 - |A(\ell, t)|^4)$.

For the short-range hopping model, $\epsilon_k^\I = -v_B\cos(k) - \mu$, we have
\begin{align}
|A^\I(\ell, t)|^2 = J_\ell (v_B t)^2 ~,
\end{align}
where $J_n(t)$ is the Bessel function of order $n$ and $v_B = v_\text{max}$.
We first consider behavior near the wavefront.
Bessel functions have so-called ``transition regions'' when the order of the Bessel function and the argument are close~\cite{olverNew1952}, here $v_B t = \ell + O(t^{1/3})$, which corresponds precisely to the wavefront region of interest to us.
In this region, we can write 
\begin{equation}
C_{nn}^\I(\ell, t) \sim f\left[\frac{(\ell - v_B t)^{3/2}}{t^{1/2}} \right]~.
\end{equation}
More precisely, the asymptotic expansion for the Bessel functions needed here is taking $\ell, v_B t$ to be very large while keeping $|\ell - v_B t|/t^{1/3}$ fixed, and it can be found in Eq.~(3.1) in Ref.~\cite{olverNew1952}.
In the regime $|\ell - v_B t|/t^{1/3} \gg 1$ this connects with the saddle-point analysis of Ref.~\cite{xuAccessing}, which gives \begin{equation}
C_{nn}^\I(\ell, t) \sim \exp[-c \frac{(\ell - v_B t)^{3/2}}{t^{1/2}}]~.    
\end{equation}
On the other hand, following the approach of Ref.~\cite{khemaniVelocitydependent}, on the fixed-velocity rays $\ell(v) = v t$ with $v > v_B$ (i.e., outside the light cone), we find $C_{nn}^\I(\ell = vt, t) \sim \exp[-\lambda(v) t]$, where $\lambda \sim(v - v_B)^{3/2}$ for small $v - v_B$; the precise asymptotic for the Bessel functions needed here is the Debye's expansion~\cite{olverNew1952} where we take $\ell, v_B t$ large while keeping $(\ell - v_B t)/t = v - v_B > 0$ fixed.
According to Ref.~\cite{olverNew1952}, the transition region's asymptotic expansion is accurate for $\ell - v_B t \ll t^{2/3}$, while the Debye's expansion is accurate for $\ell - v_B t \gg t^{1/3}$, so there is an adequate overlap between the two and hence a smooth crossover from the wavefront region to the ray region.
We remark that while we used the properties of the Bessel functions as appropriate for the specific dispersion $\epsilon(k) = -v_B \cos(k) - \mu$, the properties near the wavefront originate from behavior of $\epsilon(k)$ near the maximal group velocity, which is generic, and we expect qualitatively similar wavefront properties for any dispersion.

We also mention behavior at long times inside the light cone, $t \gg \ell/v_B$, which follows from the familiar long-time asymptotics of the Bessel functions: $C_{nn}(\ell, t)$ has oscillatory decay back to zero with envelope $\sim t^{-1}$.
Again, the long-time behavior holds also for generetic dispersion, but here it is controlled by the extrema of $\epsilon(k)$ itself.

Turning to the long-range model, we have 
\begin{equation}\label{eqn:AII(x,t)}
|A^\II(\ell, t)|^2 = \frac{2}{\pi^2} \left[ 1 - (-1)^\ell \cos(\pi v_B t) \right] \frac{(v_B t)^2}{\left[ \ell^2 - (v_Bt)^2 \right]^2} ~.
\end{equation}
The commutator function $C_{nn}(\ell, t)$ grows as $t^2$ at short time, rises sharply at the wavefront, and then decays back to zero as $t^{-2}$ inside the light cone.
Moreover, the wavefront does not broaden with time.
Indeed, consider $\ell =  v_B t + \delta \ell$, where $\delta \ell \ll v_B t$ is the small deviation from the wavefront. 
In this region, we have $C_{nn}^\II(\ell, t) \sim \frac{(v_B t)^2}{(2 v_B t + \delta \ell)^2 (\delta \ell)^2} \sim (\delta \ell)^{-2}$, which is valid when $\delta \ell$ is $O(1)$ deviation.
Comparing to the typical scaling form \cite{xuAccessing, khemaniVelocitydependent} of the wavefront $C_{nn}(\ell, t) \sim f(\delta \ell/t^\alpha)$, we have $\alpha = 0$, which formally corresponds to the absence of the wavefront broadening.
On the other hand, if we follow the rays $\ell = v t$, $v > v_B$, we have $C_{nn}^\II(\ell, t) \sim \frac{v_B^2}{(v^2 - v_B^2)^2 t^2}$, which decays as $t^{-2}$ power law at long times.
Therefore, we cannot define the velocity-dependent Lyapunov exponent here.
This is not surprising, since the Lieb-Robinson bound does not necessarily hold in this model.

We now show that in the long-range model, outside the lightcone, the early-time (perturbative) region essentially extends to the ``ray'' region---more precisely, the regime where one follows rays $\ell = vt$ with $v \gg v_B$.
Consider the long-range hopping model in terms of Pauli-matrices $X$, $Y$, and $Z$, i.e., mapping hard-core bosons spins, with $n = (1 + Z)/2$.
Introducing short-hand notation $\bar{Z}_i = -Z_i$ and the string operator $\mathcal{Z}_{i,j} = \prod_{m=i}^j \bar{Z}_m$, we write
\begin{eqnarray}\label{eqn:HII_Pauli}
H^\II &=& \frac{1}{2} \sum_{i<j} J_{ij} [X_i \mathcal{Z}_{i+1, j-1} X_j + Y_i \mathcal{Z}_{i+1, j-1} Y_j] \nonumber \\
&-& \frac{1}{2} \sum_i \mu (I_i + Z_i) ~.
\end{eqnarray}
Consider the Baker-Campbell-Hausdorff expansion of the operator $W_0(t) = \sum_{n=0}^\infty \frac{(i t)^n}{n!} L^n(W_0)$, where $L(W) \equiv [H, W]$.
The power-law growth of the commutator function is determined by the lowest-order nonzero commutator $[L^n(W_0), V_\ell]$.
Due to the long-range nature of the Hamiltonian $H^\II$, already the first order $[L(n_0), n_\ell] = \frac{1}{4}[L(Z_0), Z_\ell]$ is nonzero. 
More specifically, we have
\begin{eqnarray}
L(Z_0) &=& \sum_{j>0} J_{0j} \mathcal{Z}_{1, j-1} (-i Y_0  X_j + i X_0 Y_j) \nonumber \\
&+& \sum_{j<0} J_{j0} \mathcal{Z}_{j+1, -1} (-i X_j Y_0 + i Y_j X_0) ~,
\end{eqnarray}
giving us (assuming $\ell > 0$ for concreteness)
\begin{equation}
[L(Z_0), Z_\ell] = -2 J_{0, \ell} \mathcal{Z}_{1, \ell-1} (Y_0 Y_\ell + X_0 X_\ell) ~.
\end{equation}
The leading contribution to the commutator function is thus
\begin{equation}
C_{nn}^\II(\ell, t) \approx \frac{t^2}{32} \la|[L(Z_0), Z_\ell]|^2 \ra = \frac{(v_B t)^2}{2 \pi^2 \ell^4}[1 - (-1)^\ell] ~,
\end{equation}
where in the last equation we specialized to infinite temperature for simplicity.
For very short time $v_B t \ll 1$, this expression matches with the asymptotic behavior of the exact result, $C_{nn}(\ell, t) = \frac{1}{4} (|A(\ell, t)|^2 - |A(\ell, t)|^4) \approx \frac{1}{4} |A(\ell, t)|^2$ and using Eq.~(\ref{eqn:AII(x,t)}). 
In fact, we can also see that this asymptotic also extends ``qualitatively'' to the regime when we follow the rays $\ell = v t \gg 1$ but with $v \gg v_B$, giving us $C_{nn}^\II(\ell = vt, t) \sim v_B^2/(2 \pi^2 v^4 t^2)$ (here ``qualitatively'' means ignoring oscillations in time, which of course such Hausdorff-Campbell-Baker expansion cannot capture).

\subsection{Density-density OTOC in the Luttinger-liquid model}
\begin{figure}
\includegraphics[width=1\columnwidth]{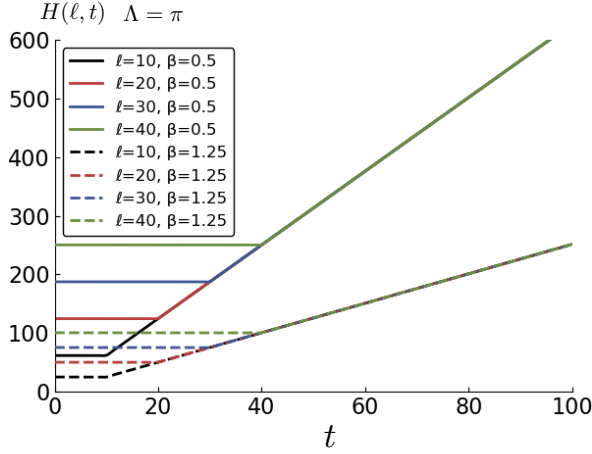}
\caption{\label{fig:Hxt}
Numerical integration results for the function $H(x,t)$ entering $C_{nn}^\III(\ell, t)$, see Eqs.~(\ref{eqn:Cnn_luttinger}) and (\ref{eqn:Hxt}), using momentum cutoff $\Lambda = \pi$.
We can see that $e^{- H(\ell, t)}$ decays exponentially inside the light cone; furthermore, for such $\ell$ and $\beta$, the numerical value is negligible compared to $1$.
}
\end{figure}

Here we present the result for the  density-density OTOC in the Luttinger liquid, while we give the detailed calculation in Appendix~\ref{app:density_luttinger}.
The non-oscillating part of this OTOC was considered in Ref.~\cite{doraOutofTimeOrdered2017}.
We consider the density operator defined in Eq.~(\ref{eqn:density_op}).
To compare with the lattice models more closely, we choose to regularize the theory by a hard-cutoff $\Lambda$, instead of a soft cutoff $e^{-\alpha |k|}$ factor in the integration over momentum $k$ used in Ref.~\cite{doraOutofTimeOrdered2017}.

The result for the commutator function is
\begin{eqnarray}\label{eqn:Cnn_luttinger}
&& C_{nn}^\III(\ell, t) = \frac{g^2}{2\pi^4} N^2(\ell, t) \\
&&~~~~~~~ + 2 d_2^4 [4 + 2\cos(4\pi\rho_0\ell) e^{-2 g H(\ell, t)}] \sin^2[2 g G(\ell, t)] \nonumber \\
&&~~~~~~~ - d_2^2 \frac{4g}{\pi^2} \cos(2\pi\rho_0\ell) e^{-2 g H(\ell, t)} N(\ell, t) \sin^2[2 g G(\ell, t)] ~, \nonumber
\end{eqnarray}
where
\begin{eqnarray}\label{eqn:Nxt}
N(\ell, t) &=& \int_0^\Lambda dk k \cos(k\ell) \sin(v_B k t) \nonumber \\
&=& \frac{\sin[\Lambda(v_B t - \ell)]}{2(v_B t - \ell)^2} - \frac{\Lambda\cos[\Lambda(v_B t - \ell)]}{2(v_B t - \ell)} \nonumber \\
&+&\frac{\sin[\Lambda(v_B t + \ell)]}{2(v_B t + \ell)^2} - \frac{\Lambda\cos[\Lambda(v_B t + \ell)]}{2(v_B t + \ell)} ~,\nonumber \\
\end{eqnarray}
\begin{eqnarray}\label{eqn:Hxt}
H(\ell, t) = \!\! \int_0^\Lambda \! \frac{dk}{k}
[2 f(v_B k) + 1] [1 - \cos(k\ell) \cos(v_B k t)] \,, ~~~~
\end{eqnarray} 
with $f(\epsilon) = 1/(e^{\beta \epsilon} - 1)$ being Bose-Einstein distribution, and
\begin{eqnarray}\label{eqn:Gxt}
G(\ell, t) &=& \int_0^\Lambda \frac{dk}{k} \cos(k\ell) \sin(v_B k t) \\
&=& \frac{1}{2} [\text{Si}(\Lambda t_+) + \text{Si}(\Lambda t_-) ]~,
\end{eqnarray}
where $\text{Si}(x) \equiv \int_0^x dy \sin(y)/y$, and we have abbreviated $t_\pm = v_B t \pm \ell$.

The above expressions are defined through the hard cutoff regularization.
(For results of the soft cutoff regularization, see Ref.~\cite{doraOutofTimeOrdered2017} and Appendix~\ref{app:density_luttinger} and \ref{app:Gxt_soft_hard}.)
First, we note that $H(\ell, t)$ is the only temperature-dependent piece; it grows linearly with $\ell$ outside the light cone, $t < \ell/v_B$, and grows linearly with $t$ inside the light cone, $t > \ell/v_B$, as shown in Fig.~\ref{fig:Hxt}. 
Therefore $e^{-g H}$ decays exponentially when either $\ell$ or $t$ are large, and can be safely neglected when discussing asymptotic behaviors.

We first consider short times, $t \ll \ell/v_B$.
It is easy to see that in this regime $N^2(\ell, t)$ grows as $t^2$.
More specifically, we have
\begin{eqnarray*}
N(\ell, t) &\sim& v_B t \int_0^\Lambda dk k^2 \cos(k\ell) \\
&=& v_B t \left(\frac{2 \Lambda \cos(\Lambda\ell)}{\ell^2} + \frac{(-2 + \ell^2\Lambda^2) \sin(\Lambda\ell)}{\ell^3} \right) ~.
\end{eqnarray*}
For the cutoff $\Lambda = \pi$ and recalling that $\ell$ is an integer, we have $N(\ell, t)^2 \sim (v_B t)^2/\ell^4$, which in fact matches with the short-time behavior of the long-range hopping model.
However, for a generic cutoff, we would obtain the leading contribution $N(\ell, t)^2 \sim (v_B t)^2/\ell^2$.

We also need to consider the contribution from $\sin^2[2 g G(\ell, t)]$ at the short time.
This can be obtained from the behavior of $G(\ell, t)$, where at the short time
\begin{eqnarray}\label{eqn:Gxt_earlytime}
G(\ell, t) &\sim& \frac{\sin(\Lambda \ell)}{\ell} v_B t \nonumber \\
&-& \frac{2 \Lambda \ell \cos(\Lambda \ell) + (-2 + \Lambda^2 \ell^2) \sin(\Lambda\ell)}{6 \ell^3}(v_B t)^3 ~.
\end{eqnarray}
For the choice $\Lambda = \pi$, we have $\sin^2[2 g G(\ell,t)] \sim 4 g^2 \pi^2 (v_B t)^6/(9\ell^4)$; while for generic $\Lambda$, we have $\sin^2[2 g G(\ell,t)] \sim 4 g^2 \sin^2(\Lambda\ell)(v_B t)^2/\ell^2$.
We therefore see that, in general, $C_{nn}^\III(\ell, t) \sim t^2$ at the short time, with the coefficient which is a function of $\ell$ whose behavior depends on the cutoff $\Lambda$, and special $\Lambda = \pi$ provides a good match with our long-range model also in the $\ell$-dependence of the coefficient.

After the wavefront passes, in the infinite time limit,
\begin{equation}
C_{nn}^\III(\ell, \infty) = 8 d_2^4 \sin^2[2 g G(\ell, \infty)] \leq 8 d_2^4 ~.
\end{equation}
For the lattice models, the commutator function is bounded by $C_{nn} \leq 2$, so we choose $d_2 = 1/\sqrt{2}$ such that the bounds match between the Luttinger liquid model and the lattice models.
Also noting that $G(\ell, \infty) = \pi/2$, we have $C_{nn}^\III \rightarrow 2 \sin^2(g\pi)$.
For the ``non-interacting'' (i.e., free-fermion)  model, $g = 1$ and $C_{nn}^\III \rightarrow 0$, which agrees with results in the lattice models.

The long-time behavior of $N^2(\ell, t)$ can be obtained easily as $N^2(\ell, t) \sim \Lambda^2\cos^2(\Lambda v_B t) \cos^2(\Lambda \ell)/(v_B t)^2$.
To analyze $\sin^2[2 g G(\ell, t)]$ at the long time, we first note the asymptotic expansion of 
\begin{eqnarray}\label{eqn:Gxt_longtime}
G(\ell, t) &\sim& \frac{\pi}{2} - \frac{\cos(\Lambda v_B t) \cos(\Lambda \ell)}{\Lambda v_B t} \nonumber \\
&-& \frac{\sin(\Lambda v_B t) \cos(\Lambda \ell)}{(\Lambda v_B t)^2} ~,
\end{eqnarray}
where we used that $\text{Si}(x) \sim \pi/2 - \cos(x)/x - \sin(x)/x^2 + O(x^{-3})$ at large $x$. 
We therefore have
\begin{eqnarray}\label{eqn:sin2gG_longtime}
\sin^2[2 g G(\ell, t)] &\sim & \sin^2(g\pi) - 2g \sin(2g\pi) \cos(\Lambda \ell) \frac{\cos(\Lambda v_B t)}{\Lambda v_B t} \nonumber \\
&+& 4g^2 \cos(2g\pi) \cos^2(\Lambda \ell) \frac{\cos^2(\Lambda v_B t)}{(\Lambda v_B t)^2} \nonumber \\
&-& 2g \sin(2g\pi) \cos(\Lambda \ell) \frac{\sin(\Lambda v_B t)}{(\Lambda v_B t)^2} ~.
\end{eqnarray}
Therefore, for the free-fermion point $g = 1$, we have that $C_{nn}^\III(\ell, t)$ vanishes as $\sim t^{-2}$ at long times.
On the other hand, if $g$ is not an integer, we have that $C_{nn}^\III(\ell, t)$ approaches a non-zero value, with the approach $\sim t^{-1}$.
The $t^{-2}$ behavior is in fact also seen in the long-range hopping model, but not in the short-range hopping model.

Lastly, we note that the wavefront does not broaden in the Luttinger liquid model, which is also the case in the long-range hopping model.
This can be seen from the fact that $C_{nn}^\III(\ell, t)$ depends on $\ell$ and $t$ only via combinations $v_B t \pm \ell$.
In this case, when one considers the behavior around the wavefront, writing $\ell = v_B t + \delta\ell$, the dependence on $\delta\ell$ has no scaling with time, which corresponds to the wavefront that does not broaden with time.
This is expected to be general feature in relativistic theories, and it is reproduced by our long-range hopping model with the straightened dispersion curve with finite band width.

\section{Boson-boson OTOC in the lattice models}\label{sec:boson_lattice}
In this section, we study the OTOC in the short-range and long-range hopping models for operators $W_0 = X_0$ and $V_\ell = X_\ell$, where $X_j \equiv b_j^\dagger + b_j$ is the combination of boson creation and annihilation operators.
[$X_j$ is simply the Pauli spin matrix $\sigma_j^x$ when the hard-core bosons are mapped to spin-1/2-s and is convenient since it is both Hermitian and unitary, see our discussion between Eqs.~(\ref{eqn:CWV}) and (\ref{eqn:FWV})].
The above operator becomes nonlocal in terms of the JW fermions. 
The calculations hence become intricate and analytical results for different asymptotic regimes are difficult to obtain.
Thus we evaluate the OTOC numerically from the full analytical expression as a Pfaffian and present results here, while we present details of the setup of calculation in Appendix~\ref{app:boson_lattice}. 
In Fig.~\ref{fig:overall}, we show the overall picture of $C_{XX}(\ell, t)$ for both the short-range and long-range hard core boson models.
We clearly observe a ballistic wavefront with butterfly velocity $v_B = 1$.
Furthermore, the commutator function saturates to a non-zero value inside the light cone, which indicates that $X_0(t)$ is evolving into a nonlocal operator, spreading throughout inside the light cone.
This behavior is in contrast to the density-density OTOC, where the commutator function goes back to zero deep inside the light cone.
In the quantum Ising model which we studied earlier in Ref.~\cite{linOutoftimeordered2018}, this type of operator that is non-local in terms of the JW fermions also shows saturation to a non-zero value in the commutator function.

Below, we examine in detail behavior of $C_{XX}(\ell, t)$ in different regimes.

\begin{figure}
\includegraphics[width=1.0\columnwidth]{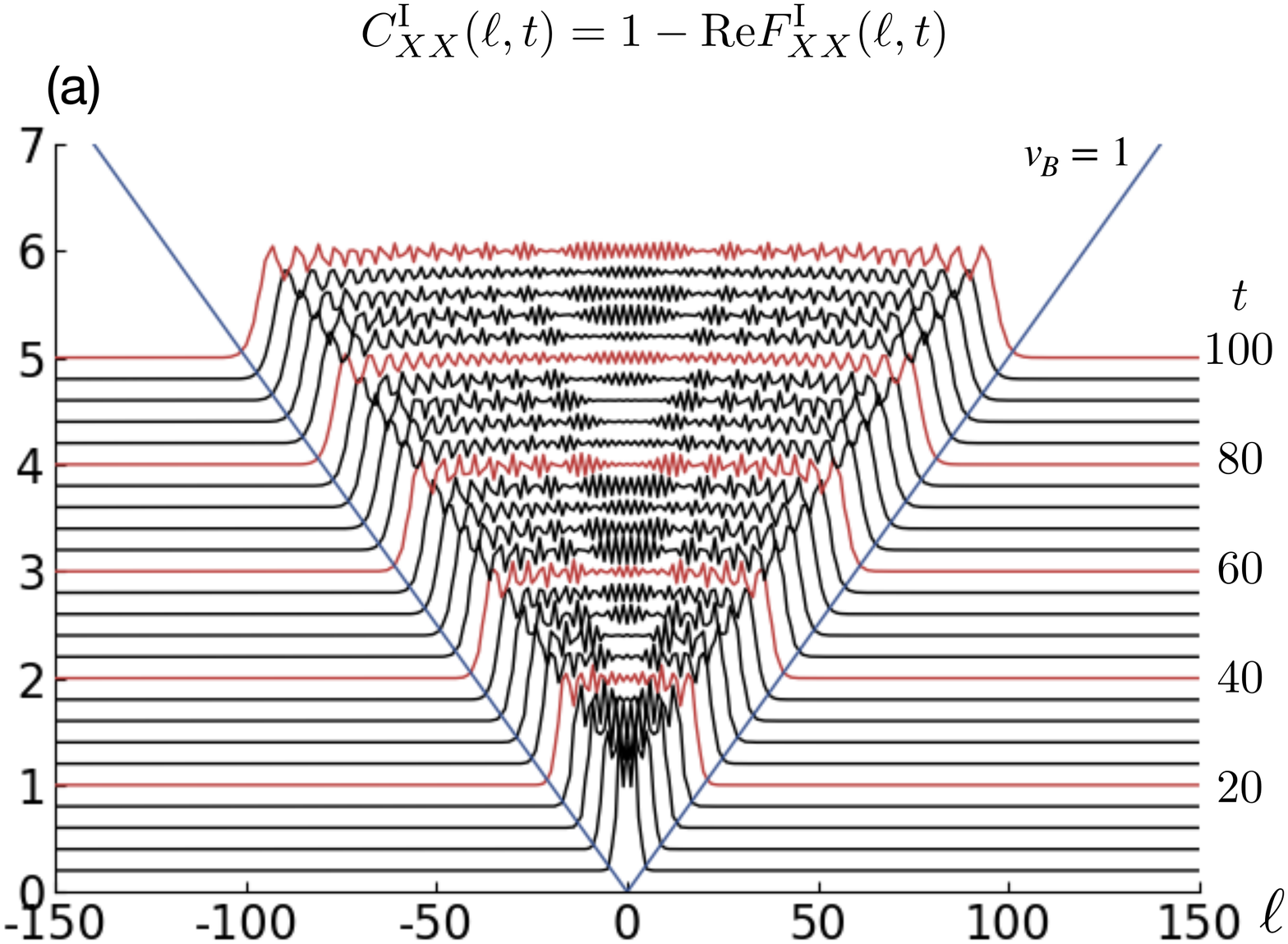}
\includegraphics[width=1.0\columnwidth]{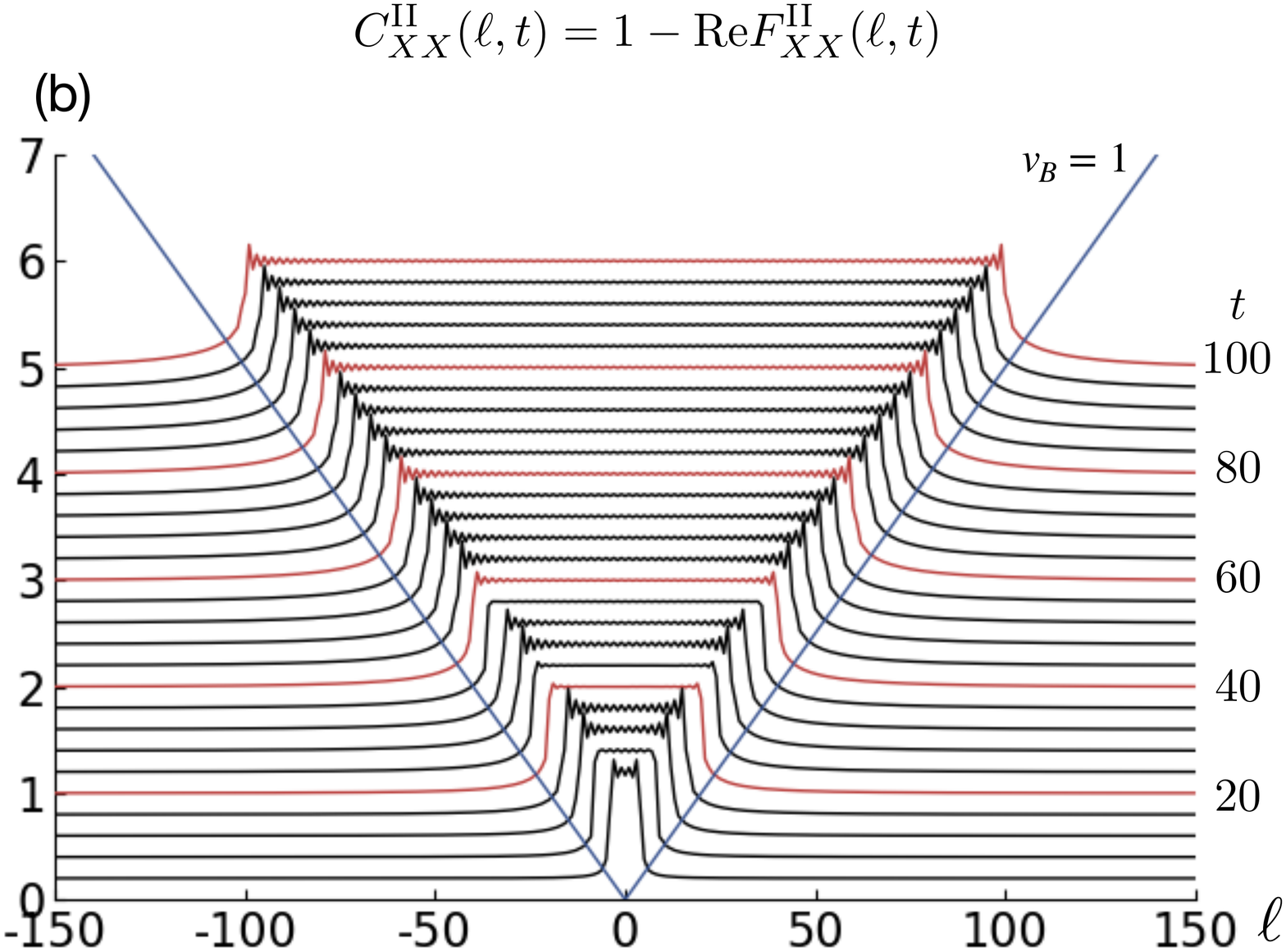}
\caption{\label{fig:overall}
The overall picture of $C_{XX}(\ell, t)$ calculated in (a) model $\I$ and (b) model $\II$, with $L = 512$, $\beta = 0$, and $J = 1$.
(At infinite temperature, the systems are at half-filling for any $\mu$.)
We have shifted each trace by $0.05 t$ to create a 3D-like visualization; also, for every time $t$ that is a multiple of $20$, we plot the trace with red color for easy reference.
The saturation to a non-zero value inside the light cone is a characteristic of scrambling; it indicates that $X_0(t)$ evolves into a nonlocal operator, in contrast to the time-evolved density operator.
}
\end{figure}

\subsection{Velocity-dependent Lyapunov exponent and wavefront broadening analysis in the short-range hopping model}
In the short-range hopping model, it is well understood that in the early-time regime, the commutator function has a position dependent power-law growth which can be understood from the Hausdorff-Campbell-Baker expansion\cite{robertsLiebRobinson2016,doraOutofTimeOrdered2017,linOutoftimeordered2018,xuAccessing} (see also Table~\ref{table:summary}).
We therefore skip the discussion of this regime in the short-range hopping model and focus on the behavior around the wavefront.

\begin{figure}
\includegraphics[width=1\columnwidth]{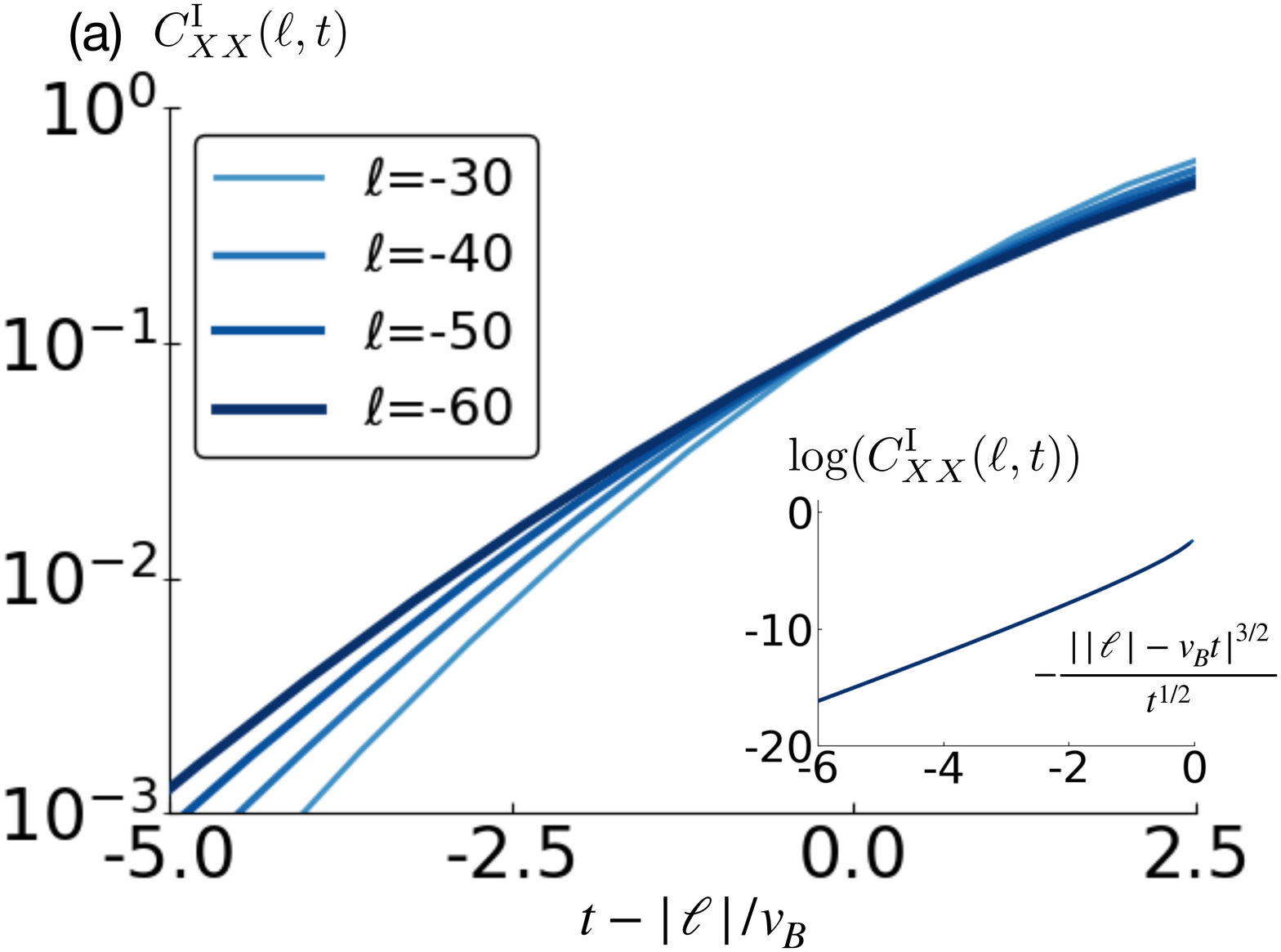}
\includegraphics[width=1\columnwidth]{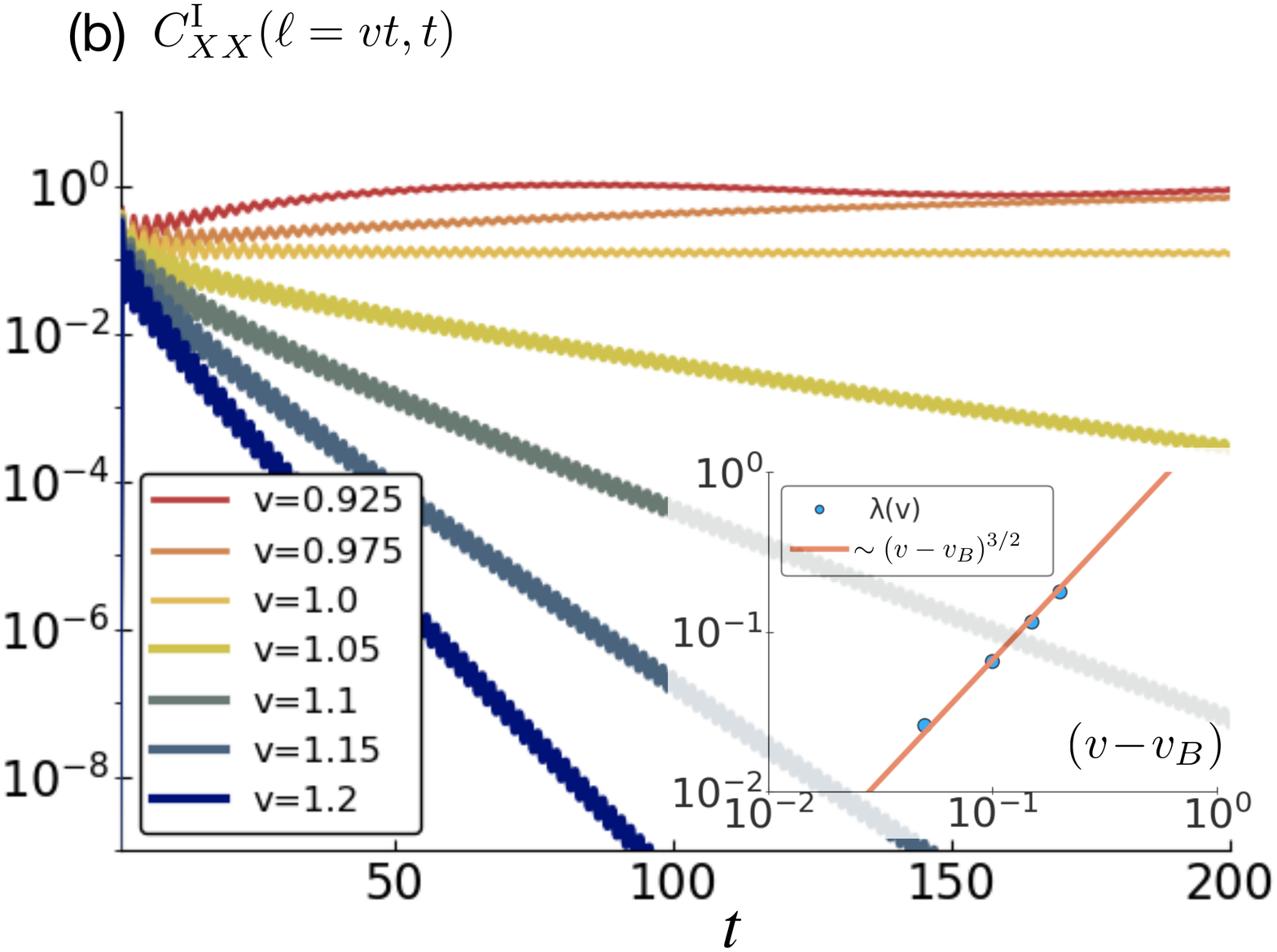}
\caption{\label{fig:short_wavefront}
(a) The commutator function $C^\I_{XX}(\ell, t)$ of the short-range model around the wave front, for several $\ell$ (negative $\ell$ correspond to points to the left of the origin in our chain with sites labeled $-L/2+1, \dots, 0, \dots, L/2$; the string that runs from the left boundary is shorter for these points than for positive $\ell$).
The systems size is $L = 512$ and the inverse temperature is $\beta = 0$.
Inset: scaling collapse demonstrating that around the wavefront, $C^I_{XX}(\ell=vt, t) \sim \exp[-\lambda (\ell - v_B t)^{3/2}/t^{1/2}]$.
(b) The commutator function $C^I_{XX}(\ell = vt, t)$ along different rays $\ell = v t$, for the same system as in panel (a).
Inset: the velocity-dependent Lyapunov exponent extracted by fitting the numerical data to $C(\ell = vt, t) = A \exp[-\lambda(v) t]$.
For velocities close to the butterfly velocity but outside the light cone, $v > v_B$, we observe the relation $\lambda(v) \sim (v - v_B)^{3/2}$.
}
\end{figure}

References~\cite{xuAccessing} and \cite{khemaniVelocitydependent} proposed that for the noninteracting free fermion models, the wavefront broadens as $t^{1/3}$.
This can be verified by either examining the long-time behavior along different fixed-velocity rays $\ell = v t$ with $v > v_B$, or by studying scaling collapse of $C(\ell, t)$ near the wavefront.
In Fig.~\ref{fig:short_wavefront}(a), we show the $t^{1/3}$ broadening by the scaling collapse analysis.
Furthermore, in Fig.~\ref{fig:short_wavefront}(b), we extracted the velocity-dependent Lyapunov exponent, which shows $(v-v_B)^{3/2}$ scaling, corresponding to the proposed $t^{1/3}$ wavefront broadening.
Note that, unlike the case of the density-density OTOC (or OTOCs composed of few fermion operators considered in Refs.~\cite{xuAccessing} and \cite{khemaniVelocitydependent}), the boson-boson OTOC does not have a simple analytical expression where the saddle-point analysis can be applied easily.
However, the wavefront broadening still has the same characteristic behavior despite the presence of the fermionic strings.
Finally, we note that our numerical results show that the above descriptions are essentially temperature independent.

\subsection{From early-time region to early-growth region in the long-range hopping model}
\begin{figure}
\includegraphics[width=1\columnwidth]{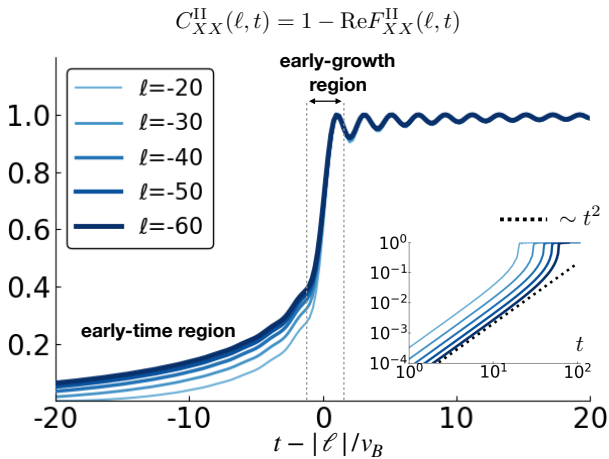}
\caption{\label{fig:long_fixell}
The commutator function $C_{XX}(\ell, t)$ in the long-range hopping model for several fixed $\ell$.
One can see that the early-time HCB region connects to the early-growth region rather abruptly. 
The window of the early-growth region always stays as $O(1)$.
Inset: demonstration of the $t^2$ power-law growth in the HCB region.}
\end{figure}

The situation with the wavefront broadening in the long-range hopping model is rather different from the short-range model, as we have already seen in the density-density OTOC.
In Fig.~\ref{fig:long_fixell}, we plot $C_{XX}^\II(\ell, t)$ for different fixed $\ell$.
In this case, as we will argue in more detail below, the perturbative (Hausdorff-Baker-Campbell) expansion gives us $t^2$ power-law growth at short time due to the long-range hopping; this is shown in the inset of Fig.~\ref{fig:long_fixell}.
Also, the early-time region connects to the early-growth region near the wavefront rather abruptly.
In fact, one can identify an $O(1)$ window around the wavefront where the $t^2$ growth stops and transits into the early-growth region.
We therefore conclude that in the long-range hopping model, the wavefront has little to no broadening.

We now provide the details of the early-time region.
Recall the Hamiltonian $H^\II$ in the spin variables written in Eq.~(\ref{eqn:HII_Pauli}). 
In the early-time (perturbative) regime, consider the expansion $X_0(t) = \sum_{n=0}^\infty \frac{(it)^n}{n!} L^n(X_0)$, where $L(A) \equiv [H, A]$.
The power-law growth of the commutator function is determined by the lowest-order nonzero commutator $[L^n(X_0), X_\ell]$.
Since $H^\II$ is long-ranged, $n = 1$ immediately ``connects'' $X_0$ and $X_\ell$, giving us $t^2$ growth.
However, there are in fact many terms that contribute to the amplitude of the $t^2$ growth.
By writing out
\begin{eqnarray}
L(X_0) &=& -i \sum_{i<0, j>0} J_{ij} \mathcal{Z}_{i+1,-1} Y_0 \mathcal{Z}_{1,j-1} (X_i X_j + Y_i Y_j) \nonumber \\
&+& i \sum_{j>0} J_{0j} \mathcal{Z}_{0,j-1} Y_j
+ i \sum_{i<0} J_{i0} Y_i \mathcal{Z}_{i+1,0} - i \mu Y_0 ~, \nonumber
\end{eqnarray}
we have (assuming $\ell > 0$ for concreteness)
\begin{eqnarray}
&&[L(X_0), X_\ell] = - 2 J_{0,\ell} \mathcal{Z}_{0,\ell} \nonumber \\
&&-2 \sum_{i<0, j>\ell} J_{ij}\mathcal{Z}_{i+1,-1} Y_0 \mathcal{Z}_{1,\ell-1} Y_\ell \mathcal{Z}_{\ell+1,j-1} (X_i  X_j +Y_i Y_j) \nonumber \\
&&+ 2 \sum_{j>\ell} J_{0j} \mathcal{Z}_{0,\ell-1} Y_\ell \mathcal{Z}_{\ell+1,j-1} Y_j
+ 2 \sum_{i<0} J_{i\ell} Y_i \mathcal{Z}_{i+1,-1} Y_0 \mathcal{Z}_{1,\ell} \nonumber ~.
\end{eqnarray}
The leading order is therefore
\begin{equation}
C_{XX}^\II(\ell, t) \sim 4 t^2 (\sum_{i<0, j>\ell} J_{ij}^2 + \sum_{j > \ell} J_{0j}^2 + J_{0,\ell}^2/2) ~,
\end{equation}
where in the thermodynamic limit $J_{ij}^2 = \frac{2 v_B^2}{\pi^2}\frac{[1 - (-1)^{i-j}]}{|i-j|^4}$.
Hence, we estimate $C_{XX}(\ell, t) \sim \frac{t^2}{\ell^2}(1 + O(\ell^{-1}))$, which is valid for $v_B t \ll 1 \ll \ell$. 
(Note that unlike the density-density OTOC, the early-time region does not extend into the ``ray'' region.)

\subsection{Long-time behavior of boson-boson OTOC}
\begin{figure}
\includegraphics[width=1\columnwidth]{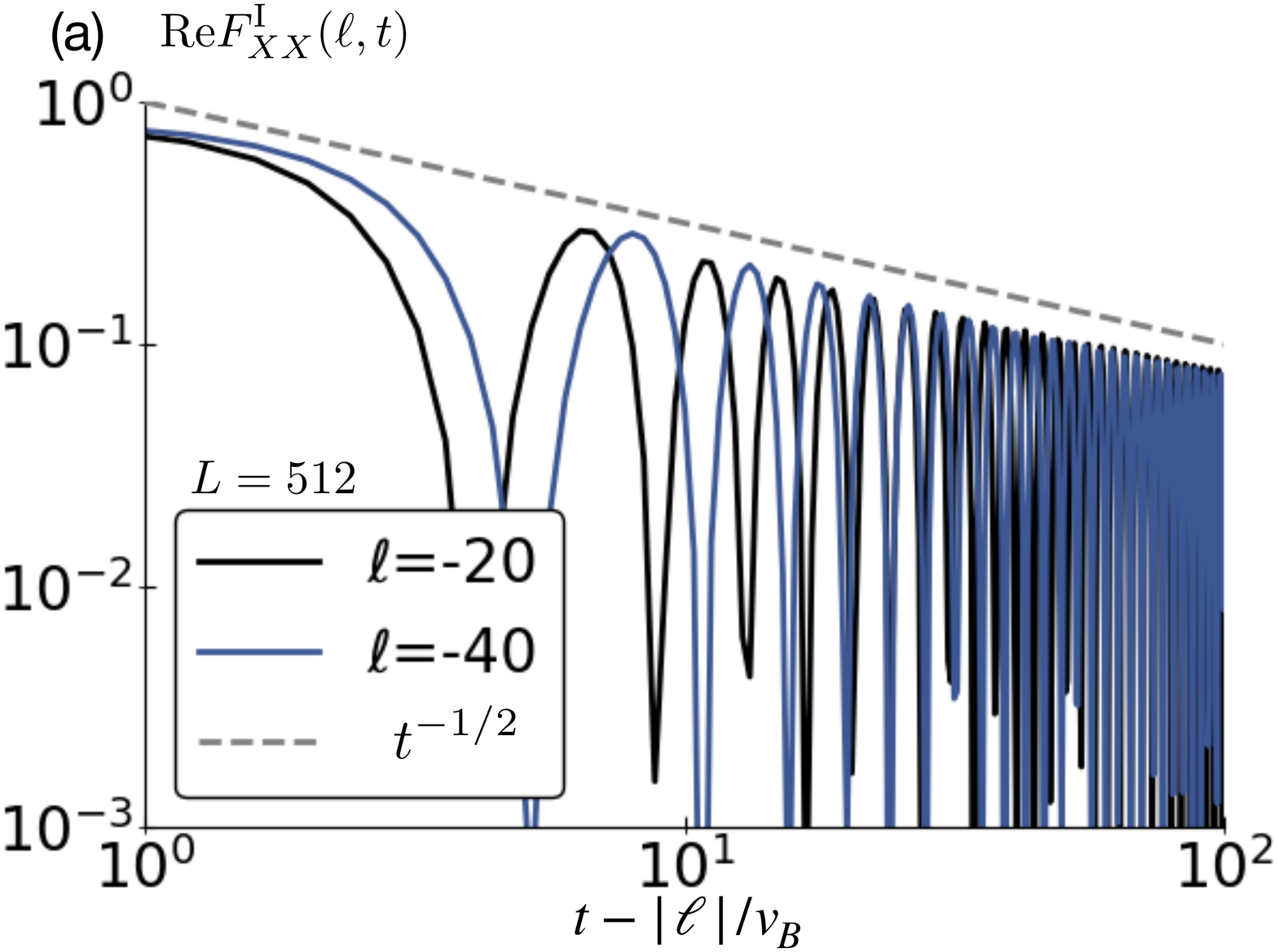}
\includegraphics[width=1\columnwidth]{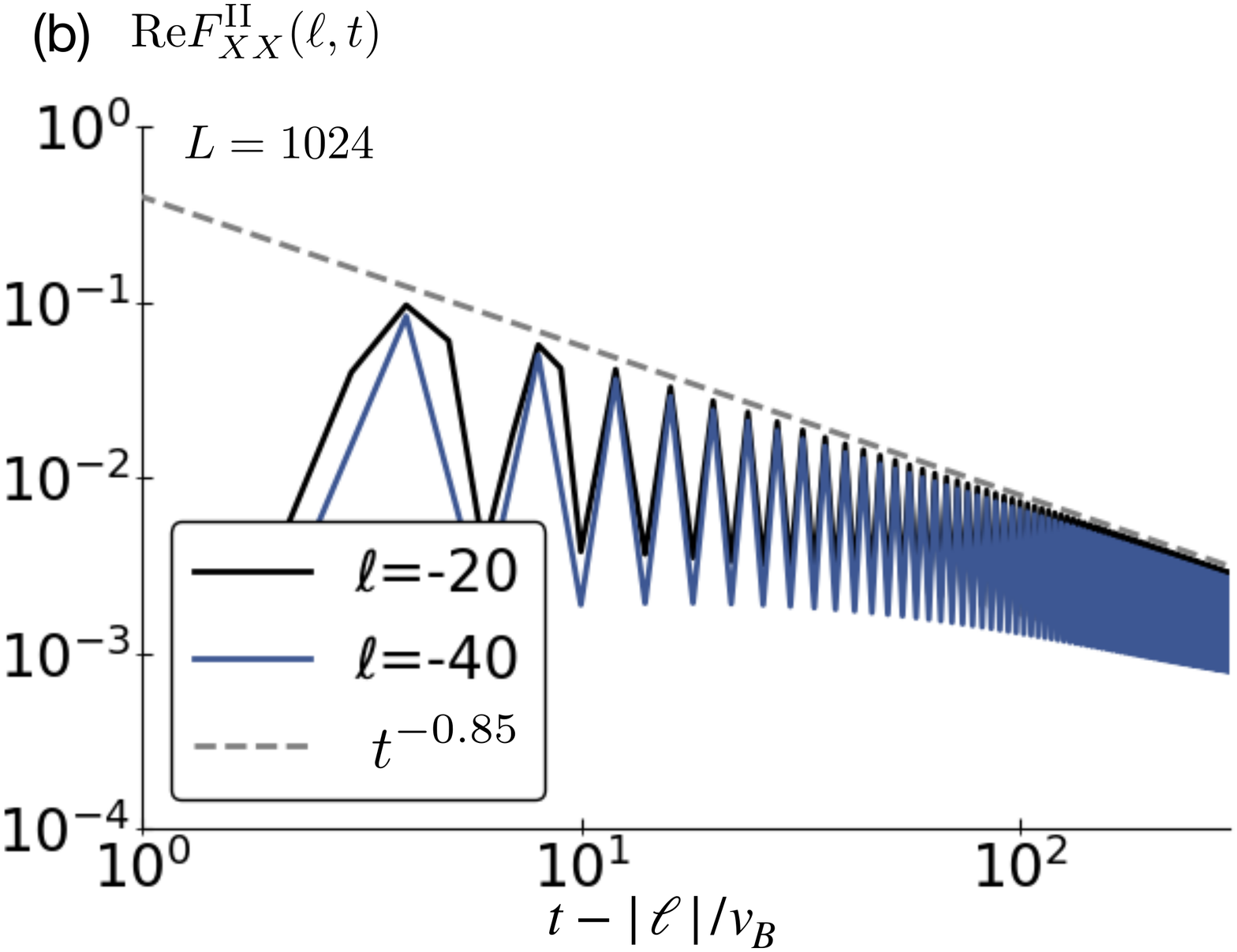}
\caption{\label{fig:lattice_longtime}
The long-time power-law decay $t^{-\alpha}$ of $\re F_{XX}(\ell, t)$ in (a) the short-range hopping model and (b) the long-range hopping model.
The numerical results suggest that the exponents are close to $\alpha^\I \approx 0.5$ and $\alpha^\II \approx 0.85$, respectively.
However, we conjecture that $\alpha^\II$ approaches $1$ in the thermodynamic limit (see the text).}
\end{figure}

Since a simple expression for $F_{XX}(\ell, t)$ is not easily obtainable, we study the long-time behavior numerically.
Figure~\ref{fig:lattice_longtime} plots the long time behavior of $F_{XX}(\ell, t)$ in the short-range hopping model and the long-range hopping model. 
It is clear that both cases show power-law decay.
For the short-range model, we conclude that the power-law is close to $t^{-1/2}$, with the system size $L = 512$ already having little finite size effect.
On the other hand, for the long-range model, we observe much stronger finite size effect (which is indeed expected).
For example, for $L = 512$, we see the power-law is close to $t^{-0.75}$, while it is close to $t^{-0.85}$ for our largest size $L = 1024$ shown in Fig.~\ref{fig:long_fixell}(b).
Extracting the power-law exponent numerically is thus challenging due to the finite-size effect, as well as due to the mixing with other power-laws and the presence of the oscillations.
We conjecture that in the thermodynamic limit, the power-law approach is $t^{-1}$, as predicted in the Luttinger liquid model, see Sec.~\ref{sec:boson_luttinger}.

To conclude, we see that the long-time power-law depends on the nature of the quasiparticle dispersion, and is different between the generic case corresponding to short-range hopping and the completely straightened case corresponding to the specific long-range hopping.  
Interestingly, in the quantum Ising case which we studied in Ref.~\cite{linOutoftimeordered2018}, similar ``chaotic'' OTOC (i.e., OTOC for operator that contains string operator in terms of the JW fermions) shows yet a different $t^{-0.25}$ power law; understanding this and the power law in the nearest-neighbor hard-core boson hopping model are outstanding questions.

\section{Boson-boson OTOC in the Luttinger liquid model}\label{sec:boson_luttinger}
In this section, we discuss the calculation of the boson-boson OTOC in the Luttinger liquid model and the agreements and disagreements with the ones obtained in the lattice models.
Here we consider the operator $X(x, t) = c (e^{i \phi(x, t)} + e^{-i \phi(x, t)})$, which resembles the sum of the boson creation and annihilation operators as considered in the previous section.
The constant $c$ will be fixed later.
We again consider the commutator function $C_{XX}(\ell, t) = \frac{1}{2} \la [X(\ell, t), X(0, 0)]^\dagger [X(\ell, t), X(0, 0)] \ra$.
To calculate it, we consider the Schwinger function in the Euclidean path integral [abbreviating $r_i = (x_i, \tau_i)$]
\begin{eqnarray}
F_{XX}^E(\{ r_i \}) &=& \la X(r_1) X(r_2) X(r_3) X(r_4) \ra \nonumber \\
&=& \sum_{p_i = \pm, \sum_i p_i = 0} \la e^{i (\sum_{i=1}^4 p_i \phi_i)} \ra  \\
&=& \sum_{p_i = \pm, \sum_i p_i = 0} \exp\left[ \frac{1}{2g} \sum_{i<j} p_i p_j K(r_i-r_j) \right] ~, \nonumber
\end{eqnarray}
where
\begin{eqnarray}
K(r) &\equiv& \int_0^\Lambda \frac{dk}{k} 2f (v_B k)[1 - \cos(kx) \cosh(v_B k \tau)]\nonumber \\
&+& \int_0^\Lambda \frac{dk}{k}[1 - \cos(kx) e^{-v_B k |\tau|}] ~,
\end{eqnarray}
with $f(\epsilon) = 1/(e^{\beta \epsilon} - 1)$ denoting the Bose-Einstein distribution function.
Again, to compare with the lattice systems at finite temperature, we choose to regularize the theory with a hard cutoff $\Lambda$. 
The $|\tau|$ symbol is to be understood as $|\tau| = \tau$ if $\re (\tau) > 0$ and $|\tau| = -\tau$ if $\re (\tau) < 0$.
To obtain the functions in the real time, we need the analytical continuations $\lim _{\epsilon \to 0^+} K(x, \tau \to \pm \epsilon + it) = H(x, t) \mp i G(x, t)$, where $H(x, t)$ and $G(x, t)$ are defined in Eqs.~(\ref{eqn:Hxt}) and (\ref{eqn:Gxt}), respectively.

Using suitable combinations of the analytical continuations, we have
\begin{equation}
C_{XX}^\III(\ell, t) = c^4 \left[8 + 4 e^{-\frac{2}{g} H(\ell, t)} \right] \sin^2 \left[\frac{G(\ell, t)}{2g} \right] ~.
\end{equation}

We can fix the constant $c$ as follows.
The factor $e^{-2 H(\ell, t)/g}$ decays exponentially to zero at long time. 
Therefore, the limiting value of the commutator function inside the light cone is $C_{XX}^{\III}(\ell, t \to \infty) = 8 c^4 \sin^2[G(\ell, \infty)/(2g)] \leq 8 c^4$. 
For the lattice models, $C_{XX}(\ell, t) \leq 2$.
We then fix $c = 1/\sqrt{2}$ so that the maximal possible value matches with the lattice models.

The integral that gives $G(x,t)$ in Eq.~(\ref{eqn:Gxt}) is convergent even if we set $\Lambda = \infty$.
In this case, $G(x,t) = \frac{\pi}{4} (\text{sign}(v_B t - x) + \text{sign}(v_B t + x)) \to \frac{\pi}{2}$ when $t \to \infty$.
This means that $C_{XX}(\ell, t) \to 1$ inside the light cone, which coincide with the result in the lattice models.
If we put the cutoff at $\Lambda = \infty$, the wavefront becomes a step function, and the long time behavior is described by $\exp[-2 H(x,t)/g]$ (which in fact approaches zero when $\Lambda \to \infty$).
A more realistic approach (i.e., closer to the lattice models) is to have a finite cutoff $\Lambda$.
As we will see, introducing the finite cutoff indeed changes the behavior $C_{XX}(\ell, t)$ around the wavefront and its asymptote in the long time.

\begin{figure}
\includegraphics[width=1\columnwidth]{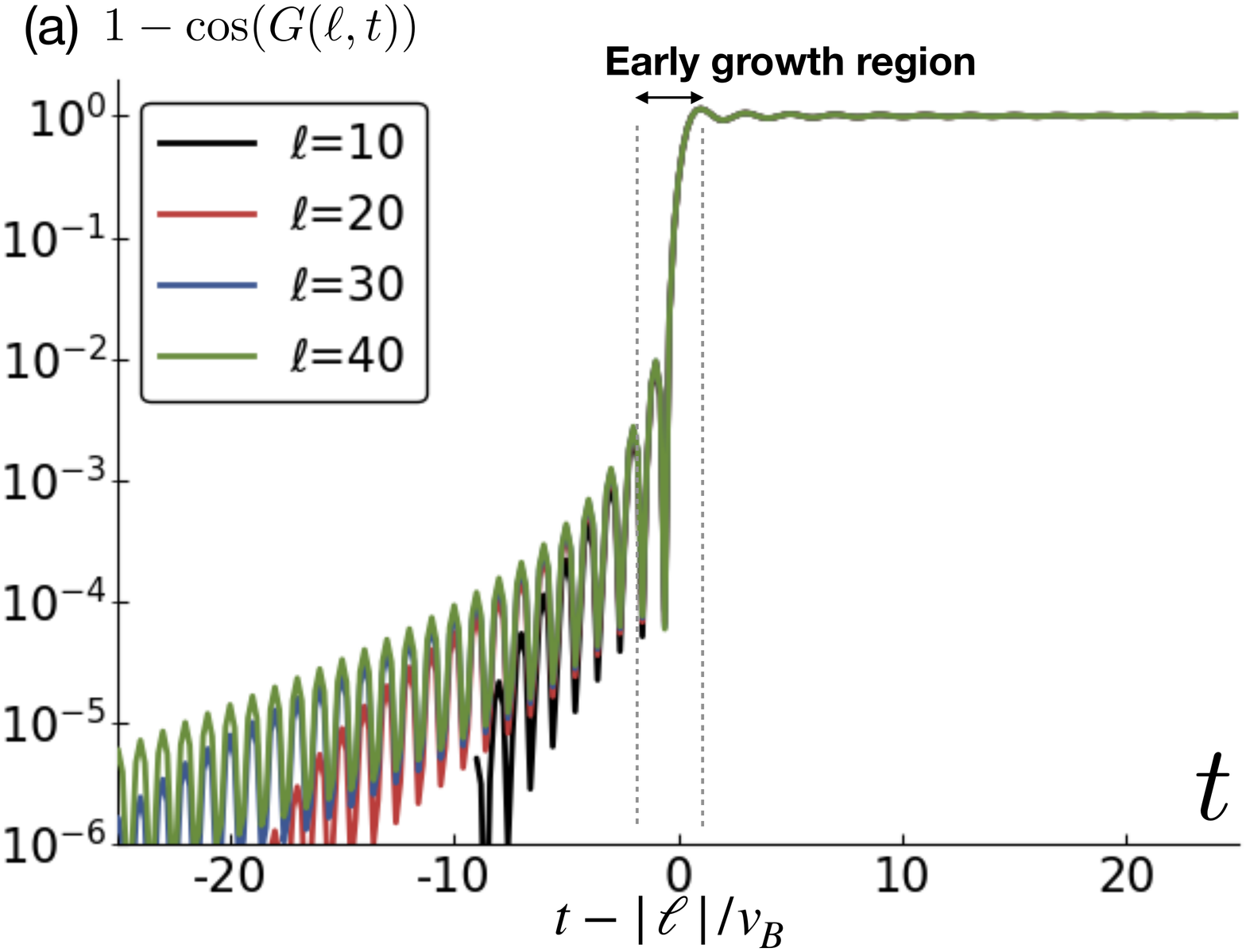}
\includegraphics[width=1\columnwidth]{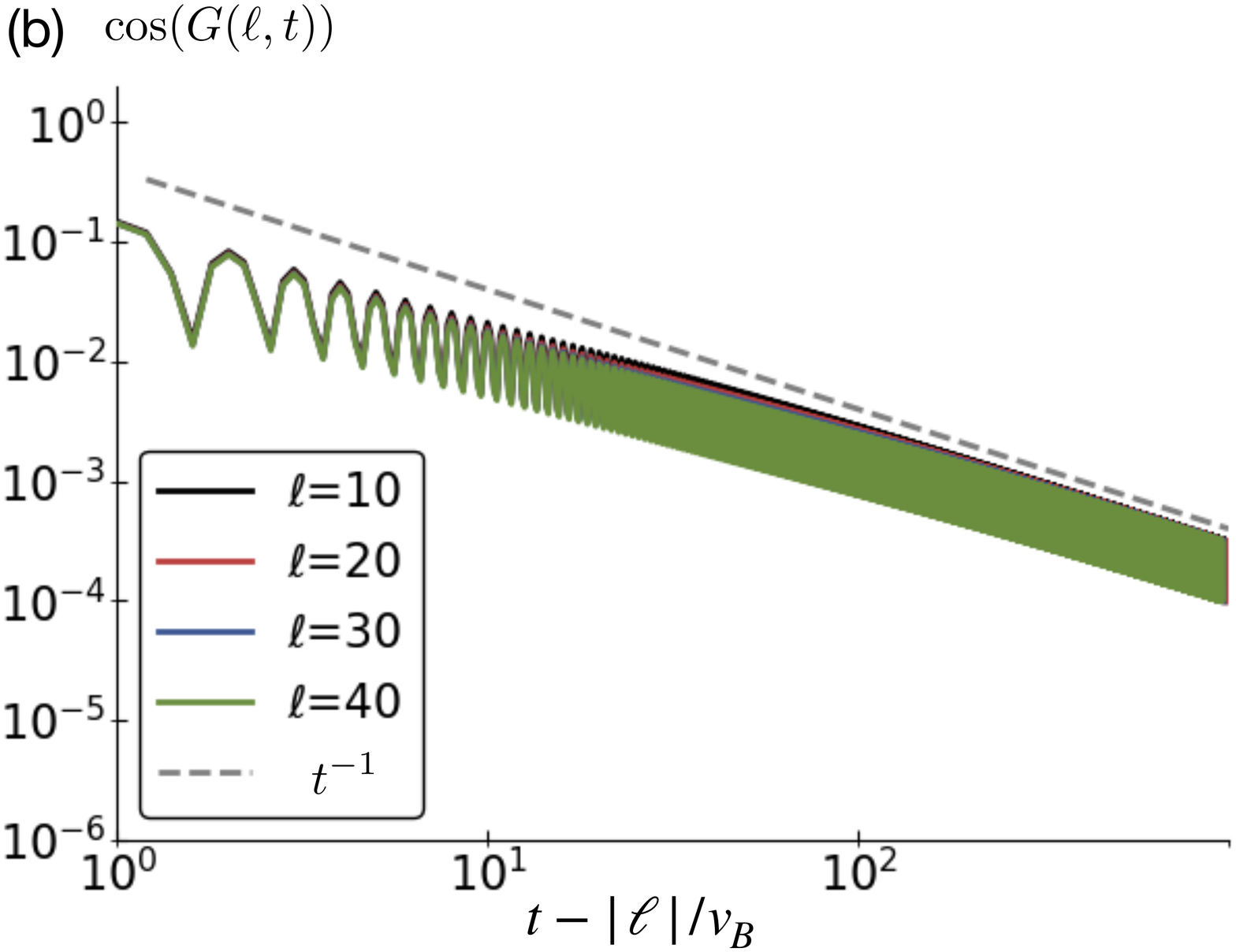}
\caption{\label{fig:Lutt_boson_otoc}
(a) The dominant part of the commutator function $C_{XX}^\III(\ell, t) \approx 1 - \cos[G(\ell, t)]$ in the Luttinger model with $g = 1$ (``free-fermion'' value).
(b) The long-time approach is described by the $t^{-1}$ power law.}
\end{figure}

Figure~\ref{fig:Lutt_boson_otoc} shows results with finite cutoff $\Lambda = \pi$ for the case with the Luttinger parameter $g = 1$ (which corresponds to non-interacting fermions). 
In the figure, we ignored the $\exp[-2 H(\ell, t)/g]$ since it is numerically negligible and does not affect the behavior of $C_{XX}(\ell, t)$, as explained in Fig.~\ref{fig:Hxt}.
One can see that the presence of the finite cutoff indeed modifies the shape of the wavefront (early-growth region), which is no longer the simple step function.
However, the time window of the early-growth is always $O(1)$, which is indeed similar to the case of the lattice model with long-range hopping.
Moreover, in this case as well the wavefront does not broaden, which can be also seen from the fact that $C_{XX}^\III(\ell, t)$ is a function of $v_B t \pm \ell$ only.
Making a more detailed comparison of Fig.~\ref{fig:Lutt_boson_otoc}(a) and 
\begin{equation}
G(\ell, t) \approx \pi/4 + \text{Si}[\Lambda (v_B t - \ell)]/2
\label{eq:G_LL_wavefront}
\end{equation}
in this early-growth region, $v_B t \sim \ell \gg 1$, we note that the shape of the wavefront is indeed converging as $\ell$ increases.
The precipitous drop when going away from the wavefront for $v_B t < \ell$ but still in the early growth region is due to numerical ``accident:'' here $G(\ell, t)$ oscillates around $0$, hence the strong drop in $1 - \cos[G(\ell,t)]$, and the next largest value happens to be small by accident.

Before the wavefront is reached, $C_{XX}(\ell, t)$ has a fixed power-law growth $\sim t^2$ for any $\ell$, which is also similar to the early-time (perturbative) region in the long-ranged hopping hard-core boson model.
To be more specific, recalling the early time behavior of $G(\ell, t)$ in Eq.~(\ref{eqn:Gxt_earlytime}), similarly we have $\sin^2[G(\ell, t)/(2g)] \sim \pi^2 (v_B t)^6/(36 g^2 \ell^4)$ for the choice $\Lambda = \pi$; while for a generic $\Lambda$, we have $\sin^2[G(\ell, t)/(2g)] \sim \sin^2(\Lambda\ell) (v_B t)^2/(4 g^2 \ell^2)$.
Note that in this case, one has to choose generic cut-off to match the early-time growth power law in the long-range hopping model.

We note that the long-time power-law approach is also due to the presence of the finite cutoff.
Recalling the long-time behavior of $G(\ell, t)$ in Eq.~(\ref{eqn:Gxt_longtime}), we have
\begin{eqnarray}
\sin^2\left[\frac{G(\ell, t)}{2g} \right] &\sim& \sin^2\left(\frac{\pi}{4g}\right) -  \sin\left(\frac{\pi}{2g} \right)\cos(\Lambda \ell) \frac{\cos(\Lambda v_B t)}{2g(\Lambda v_B t)} \nonumber \\
&+&  \cos\left(\frac{\pi}{2g} \right)\cos^2(\Lambda \ell) \frac{\cos^2(\Lambda v_B t)}{4g^2(\Lambda v_B t)^2} \nonumber \\
&-&  \sin\left(\frac{\pi}{2g} \right) \cos(\Lambda \ell)\frac{\sin(\Lambda v_B t)}{2g(\Lambda v_B t)^2} ~.
\end{eqnarray}
Therefore, for general $g \neq 1/(2m)$, where $m$ is some integer, we have $C_{XX}^\III(\ell, t) \sim t^{-1}$ at long time; this includes the non-interacting model $g = 1$ [in the special cases with $g = 1/(2m), m \in \mathbb{Z}$, we have $C_{XX}^\III(\ell, t) \sim t^{-2}$].
We speculate that this agrees with the long-range hard-core boson model in the thermodynamic limit.
We therefore see that, while not in all the details, the Luttinger liquid model can capture a great deal of the OTOC behavior in the long-range hopping hardcore boson model.

\section{Conclusions}\label{sec:conclusion}
In this paper, we studied the OTOCs in the hard-core boson models with short-range hopping and with long-range hopping where we artificially straighten the fermionic quasiparticle dispersion.
We compared these models to the Luttinger-liquid model with hard cutoff regularization, which mimics the finite band width in the lattice models.

The density-density commutator function exhibits ``non-scrambling'' behavior (i.e., it approaches zero in the long-time limit) in all three models.
In the short-range hopping model, the wavefront broadens as $t^{1/3}$, which can be verified using the asymptotic properties of the Bessel functions in the so-called ``transition region.''
On the other hand, in the long-range hopping model and the Luttinger-liquid model, we find that the wavefront does not broaden; there is also no well-defined ``exponential growth'' (i.e., ``butterfly effect'') regime since the wavefront width is finite (and the width is also cutoff-dependent in the Luttinger-liquid model).

Before the wavefront reaches, the long-range model and the Luttinger-liquid model both show $t^2$ growth.
The $t^2$ growth in the long-range model can be understood using the perturbative early-time expansion and is due to the fact that all sites ``talk'' to each other via the long-range couplings.
The coefficient of the $t^2$ growth as a function of $\ell$ can be carried out according to the perturbative expansion.
The fact that we find similar early-time behavior in the Luttinger-liquid model suggests that it should be regarded as representing bosons with long-range couplings.
We therefore see that this ``light-cone leakage" phenomenon\cite{luitzEmergent} might be a very general feature for long-range models.
The short-range model, on the other hand, shows position-dependent power-law growth in the early-time regime described by the perturbative expansion.
After the wavefront passes, both the long-range model and the Luttinger-liquid model show $t^{-2}$ decay, while the short-range model shows $t^{-1}$ decay.

Turning to the boson-boson commutator function, the calculations are more complex in the lattice models (since the boson operator contains a string operator when expressed in terms of the JW fermions) and require numerical calculations, while they are still analytically tractable in the Luttinger-liquid model.
We find that the boson-boson commutator function shows saturation inside the light cone in all free models.
Such a characteristic ``scrambling'' behavior reflects the fact that the boson operator turns into a highly nonlocal operator under the Heisenberg evolution. 
In the short-range hopping model, we found the $t^{1/3}$ wavefront broadening by wavefront scaling collapse and by extracting the velocity-dependent Lyapunov exponent.
In the long-range hopping model, we find a nonbroadening wavefront, which is also the case in the Luttinger-liquid model;
in both cases, one cannot define a parametrically large window to describe the wavefront behavior asymptotically [while the sharp onset in the Luttinger-liquid model in Fig.~\ref{fig:Lutt_boson_otoc} is reminiscent of an exponentially growing wavefront, we emphasize that this is a numerical accident for the simple function describing the wavefront at long times, Eq.~(\ref{eq:G_LL_wavefront})].
As far as the wavefront broadening is concerned, in all free models, the boson-boson and density-density OTOCs thus show similar broadening behavior.
After the wavefront passes, the boson-boson OTOC approaches its limiting value as $t^{-1}$ in the long-range model and the Luttinger-liquid model, while the approach is $t^{-0.5}$ in the short-range model.

We thus see that, despite the integrability of the models, different operators can still show different behaviors in the OTOC.
It was argued that in rational conformal theories, the $t = \infty$ values of the OTOCs are solely determined by the topological data associated with the operators in the models~\cite{guFractional2016}.
While we do not expect the full conformal symmetry in the lattice models at finite temperature, the topological data might still be present, resulting in the same $t = \infty$ values of the OTOCs independent of the details of the dynamics.
On the other hand, the character of the wavefront broadening and the long-time power-law approach depends on the details of the dispersion relations of the quasiparticles.
In particular, one lesson from our study is that the conformal field theories cannot be used to described such properties of short-range models at finite temperatures.

Lastly, we mention some outstanding questions.
While a seemingly simple description appears to exist for the boson-boson OTOC results, we could not acquire a more analytical understanding due to its intricacy.
Thus, it will indeed be valuable if one can obtain some analytical understanding regarding the long-time asymptotic or the wavefront behavior, which we only obtained numerically.
We have also shown the feasibility of reconciling the OTOCs of the hard-core boson model with a completely straightened quasiparticle dispersion to the OTOCs in the linear-Luttinger liquid.
It may therefore be possible to match the behavior of OTOCs of the short-range boson model and of a ``nonlinear'' Luttinger liquid (i.e., theory that includes some ``band curvature'' effects)~\cite{imambekovOnedimensional2012}, which is worth pursuing.
The quasiparticle description behind the systems we studied in this paper are noninteracting fermions.
Another question is how the details of the OTOC change when one adds interactions, and what role the integrability plays.
A recent work~\cite{gopalakrishnanOperator2018a} has shown that in an interacting integrable Floquet system, the OTOCs have diffusive wavefront broadening, similar to the random unitary circuit model.
On the other hand, little is known for integrable Hamiltonian systems that do not have a description in terms of free particles (e.g., models where the Jordan-Wigner fermions are interacting).
A robust study on such systems such as the XXZ chain will be valuable for a deeper understanding of OTOCs and operator spreading in high-energy, quantum-information, and condensed-matter communities.

\begin{acknowledgments}
The authors would like to thank Y.-Z.~Chou, D.~Huse, N.~Hunter-Jones, V.~Khemani, B.~Swingle, and N.~Yunger Halpern for useful discussions.
This work was supported by NSF through Grant No. DMR-1619696, and also by the Institute for Quantum Information and Matter, an NSF Physics Frontiers Center, with support of the Gordon and Betty Moore Foundation.
\end{acknowledgments}

\appendix
\section{Calculation of the density-density OTOC in the lattice models}\label{app:density-density_lattice}
Here we present detailed calculations for the density-density OTOC in the short-range and long-range hopping models.
Directly calculating the ``commutator-squared'' $|[n_0(t), n_\ell]|^2$ using the Hamiltonian formalism and the explicit Heisenberg evolution of the operators is in fact easier than calculating the expanded four terms individually.
However, here we will carry out the calculations using path-integral formalism and obtain all the terms individually.
Such approach parallels OTOC calculations in field theories, and the Luttinger liquid model is one example that we want to compare and contrast.

We define
\begin{eqnarray*}
C_{nn}(\ell, t) &=& \frac{1}{2} \la [n_0(t), n_\ell]^\dagger [n_0(t), n_\ell] \ra \\
&=& \frac{1}{2}\left[F_1(\ell, t) + F_2(\ell, t) - F_3(\ell, t) - F_4(\ell, t) \right] ~,
\end{eqnarray*}
where
\begin{eqnarray*}
F_1(\ell, t) &=& \la n_\ell \, n_0(t) \, n_0(t) \, n_\ell \ra ~, \\
F_2(\ell, t) &=& \la n_0(t) \, n_\ell \, n_\ell \, n_0(t) \ra ~, \\
F_3(\ell, t) &=& \la n_0(t) \, n_\ell \, n_0(t) \, n_\ell \ra ~, \\
F_4(\ell, t) &=& \la n_\ell \, n_0(t) \, n_\ell \, n_0(t) \ra ~.
\end{eqnarray*}

Now consider the Jordan-Wigner fermion model formulated in the Euclidean path integral over Grassmann fields, $Z = \int \pathD[\bar{\eta}, \eta] e^{-S[\bar{\eta},\eta]}$, where
\begin{equation}
S = \int_0^\beta d\tau \sum_{ij} \bar{\eta}_i [\delta_{ij} \partial_\tau  + J_{ij}] \eta_j ~.
\end{equation}
The standard method of calculating OTOC is to calculate the Schwinger functions and do the analytical continuation to the Wightmann functions. 
Consider the Schwinger function
\begin{equation}
F^{(E)}(\ell; \tau_1, \tau_2, \tau_3, \tau_4) = \la T_\tau \{ n_\ell(\tau_1) n_\ell(\tau_2) n_0(\tau_3) n_0(\tau_4) \} \ra ~.
\end{equation}
The required Wightmann functions are obtained by 
\begin{eqnarray*}
F_1(\ell, t) &=& F^{(E)}(\ell; \epsilon_1, \epsilon_4, \epsilon_2 + it, \epsilon_3 + it) ~, \\
F_2(\ell, t) &=& F^{(E)}(\ell; \epsilon_2, \epsilon_3, \epsilon_1 + it, \epsilon_4 + it) ~, \\
F_3(\ell, t) &=& F^{(E)}(\ell; \epsilon_2, \epsilon_4, \epsilon_1 + it, \epsilon_3 + it) ~, \\
F_4(\ell, t) &=& F^{(E)}(\ell; \epsilon_1, \epsilon_3, \epsilon_2 + it, \epsilon_4 + it) ~,
\end{eqnarray*}
where the $\epsilon_i$'s are taken to the limit of $0^+$ in the order of $\epsilon_1 > \epsilon_2 > \epsilon_3 > \epsilon_4 > 0$.

The Schwinger function is in fact an eight-point fermion correlation function
\begin{widetext}
\begin{eqnarray}
F^{(E)}(\ell; \tau_1, \tau_2, \tau_3, \tau_4) &=& \la \eta_\ell(\tau_1)\eta_\ell(\tau_2) \eta_0(\tau_3) \eta_0(\tau_4) \bar{\eta}_\ell(\tau_1+\delta) \bar{\eta}_\ell(\tau_2+\delta) \bar{\eta}_0(\tau_3+\delta)\bar{\eta}_0(\tau_4+\delta) \ra \\
&=&\det 
\begin{bmatrix}
\la n_\ell \ra & \la \eta_\ell(\tau_1) \bar{\eta}_\ell(\tau_2) \ra & \la \eta_\ell(\tau_1) \bar{\eta}_0(\tau_3) \ra & \la \eta_\ell(\tau_1)\bar{\eta}_0(\tau_4) \ra \\
\la \eta_{\ell}(\tau_2)\bar{\eta}_\ell(\tau_1)\ra & \la n_\ell\ra & \la \eta_{\ell}(\tau_2)\bar{\eta}_0(\tau_3)\ra & \la \eta_{\ell}(\tau_2)\bar{\eta}_0(\tau_4)\ra \\
\la \eta_{0}(\tau_3)\bar{\eta}_\ell(\tau_1)\ra & \la \eta_{0}(\tau_3)\bar{\eta}_\ell(\tau_2)\ra & \la n_0\ra & \la \eta_{0}(\tau_3)\bar{\eta}_0(\tau_4)\ra \\
\la \eta_{0}(\tau_4)\bar{\eta}_\ell(\tau_1)\ra & \la \eta_{0}(\tau_4)\bar{\eta}_\ell(\tau_2)\ra & \la \eta_{0}(\tau_4)\bar{\eta}_0(\tau_3)\ra & \la n_0\ra 
\end{bmatrix} ~.
\end{eqnarray}
\end{widetext}
Note that in the first line, $\delta$ is a positive infinitesimal smaller than all $\epsilon_i$, i.e., $\delta < \epsilon_i$, $i=1, \dots, 4$; in the second line, $\delta$ has been taken to be $0^+$. 

We can now obtain the Wightmann functions.
We have
\begin{eqnarray}
F_1&=&\det
\begin{bmatrix}
\la n_\ell\ra & \la n_\ell\ra-1 & -\la c_{\ell}c_0^{\dagger}(t)\ra & -\la c_{\ell}c_0^{\dagger}(t)\ra \\
\la n_\ell\ra & \la n_\ell\ra & \la c_{0}^{\dagger}(t)c_\ell\ra & \la c_{0}^{\dagger}(t)c_\ell\ra \\
\la c_{\ell}^{\dagger}c_0(t)\ra & -\la c_{0}(t)c_\ell^{\dagger}\ra & \la n_0\ra & \la n_0\ra-1 \\
\la c_{\ell}^{\dagger}c_0(t)\ra & -\la c_{0}(t)c_\ell^{\dagger}\ra & \la n_0\ra & \la n_0\ra  
\end{bmatrix}~, \nonumber \\
&=&\la n_\ell \ra \la n_0 \ra +\la n_\ell \ra |A(\ell,t)|^2-|\la c_0(t)^{\dagger}c_\ell \ra|^2~,
\end{eqnarray}
where $A(\ell, t) = \la c_\ell^\dagger c_0(t) + c_0(t) c_\ell^\dagger\ra = \frac{1}{L} \sum_k e^{ik\ell - i\epsilon(k)t}$ is just the fermion evolution function. 
Note that it is completely independent of the temperature.

Next, we have 
\begin{eqnarray}
F_2&=&\det
\begin{bmatrix}
\la n_\ell\ra & \la n_\ell\ra-1 & \la c_{0}^{\dagger}(t)c_\ell\ra & -\la c_{\ell}c_0^{\dagger}(t)\ra \\
\la n_\ell\ra & \la n_\ell\ra & \la c_{0}^{\dagger}(t)c_\ell\ra & -\la c_{\ell}c_0^{\dagger}(t)\ra \\
-\la c_{0}(t)c_\ell^{\dagger}\ra & -\la c_{0}(t)c_\ell^{\dagger}\ra & \la n_0\ra & \la n_0\ra-1 \\
\la c_{\ell}^{\dagger}c_0(t)\ra & \la c_{\ell}^{\dagger}c_0(t)\ra & \la n_0\ra & \la n_0\ra  
\end{bmatrix}~, \nonumber \\
&=&\la n_\ell \ra \la n_0 \ra +\la n_0 \ra |A(\ell,t)|^2-|\la c_0(t)^{\dagger}c_\ell \ra|^2~,
\end{eqnarray}
and
\begin{eqnarray}
F_3&=&\det
\begin{bmatrix}
\la n_\ell\ra & \la n_\ell\ra-1 & \la c_{0}^{\dagger}(t)c_\ell\ra & -\la c_{\ell}c_0^{\dagger}(t)\ra \\
\la n_\ell\ra & \la n_\ell\ra & \la c_{0}^{\dagger}(t)c_\ell\ra & \la c_{0}^{\dagger}(t)c_\ell\ra \\
-\la c_{0}(t)c_\ell^{\dagger}\ra & -\la c_{0}(t)c_\ell^{\dagger}\ra & \la n_0\ra & \la n_0\ra-1 \\
\la c_{\ell}^{\dagger}c_0(t)\ra & -\la c_{0}(t)c_\ell^{\dagger}\ra & \la n_0\ra & \la n_0\ra
\end{bmatrix}~, \nonumber \\
&=&(|A(\ell,t)|^2+1)(\la n_\ell \ra\la n_0 \ra+\la c_0^{\dagger}(t)c_\ell\ra \la c_0(t) c_\ell^{\dagger} \ra) \nonumber \\
&-&A(\ell,t)\la c_0^{\dagger}(t)c_\ell\ra~.
\end{eqnarray}
Finally, we use 
\begin{eqnarray}
F_4=F_3^{*}&=&(|A(\ell,t)|^2+1)(\la n_\ell \ra\la n_0 \ra+\la c_\ell^{\dagger}c_0(t)\ra \la c_\ell c_0^{\dagger}(t)  \ra) \nonumber \\
&-&A^{*}(\ell,t)\la c_\ell^{\dagger}c_0(t)\ra~.
\end{eqnarray}
Combining everything, we have 
\begin{eqnarray}
C_{nn}(\ell,t)&=&|A(\ell,t)|^2 \Big\{ (\la n_\ell\ra+\la n_0\ra)/2-\la n_\ell\ra\la n_0\ra \nonumber \\
&& ~~~~~~~~~~~~ -\re[\la c_0^{\dagger}(t)c_\ell\ra \la c_0(t) c_\ell^{\dagger} \ra] \Big\} ~.
\end{eqnarray}
It is easy to see that the most important part of $C_{nn}(\ell, t)$ is the temperature-independent factor $|A(\ell, t)|^2$.
Indeed, the first line in the $\{ \dots \}$ is a non-zero constant [equal to $d(1-d)$ where $d$ is the density], while the second line decays both in separation $\ell$ and in time $t$.

\section{Calculation of the density-density OTOC in the Luttinger liquid model}\label{app:density_luttinger}
In this section, we present detailed calculations for the density-density OTOC. 
The non-oscillation component of the density-density OTOC was calculated in Ref.~\cite{doraOutofTimeOrdered2017}. 
In fact, it is the dominant contribution for $C_{nn}(\ell,t)$. 
Here, for the completeness, we also include the $q = 2k_F$ component of $C_{nn}(\ell, t)$ and use the path-integral formalism instead of the Hamiltonian formalism.

The density operator in consideration is defined in Sec.~\ref{sec:models}. 
Here we repeat the definition for reader's convenience.
\begin{equation}
n(x) = d_0 + \rho_0(x) + d_2 W(x) ~,
\end{equation}
where $\rho_0(x) \equiv -\partial_x \hat{\theta}(x)/\pi$ and 
\begin{equation}
W(x) \equiv e^{i 2\pi d_0 x} V_{-2}(x) + e^{-i 2\pi d_0 x} V_2(x) ~,
\end{equation}
with the abbreviation $V_m(x) \equiv e^{i m \hat{\theta}(x)}$, $d_0 = k_F/\pi$ denoting the density, and $d_2$ is some constant determined by microscopic details of the model.

We therefore have
\begin{eqnarray}
C_{nn}(\ell, t) &=& C_{\rho_0\rho_0}(\ell, t) + d_2^4 C_{WW}(\ell, t) \nonumber \\
&+& d_2^2 C_{\rho_0 W}(\ell, t) + d_2^2 C_{W \rho_0}(\ell, t) \nonumber \\
&+& d_2^2 C_{\rho_0 \rho_0 W W}(\ell, t) + d_2^2 C_{\rho_0 W W \rho_0}(\ell, t) ~,
\end{eqnarray}
where we have abbreviated
\begin{eqnarray*}
C_{\rho_0 \rho_0 W W}(\ell, t) &\equiv& \frac{1}{2} \la [\rho_0(x,t),\rho_0(0)]^\dagger [W(x, t), W(0)] \ra + \hc ~, \\
C_{\rho_0 W W \rho_0}(\ell, t) &\equiv& \frac{1}{2} \la [\rho_0(x, t), W(0)]^\dagger [W(x, t), \rho_0(0)] \ra + \hc ~.
\end{eqnarray*}
To calculate the various commutator functions, we first consider the Schwinger functions and then analytically continue to the desired combinations.

First, we consider $C_{\rho_0 \rho_0}(\ell, t)$.
Abbreviating $r \equiv (x, \tau)$, we define (with the hard-cutoff $\Lambda$)
\begin{eqnarray}
D(r)&=&\int_0^{\Lambda}dkk2f_B(v_Bk)\cosh(v_Bk\tau)\cos(kx) \nonumber \\
&+&\int_0^{\Lambda}dkk\cos(kx)e^{-|\tau|vk}~.
\end{eqnarray}
(For the soft-cutoff version, one integrates $k$ from $0$ to $\infty$ with an extra factor $e^{-\alpha k}$.)
We in fact have $\partial_{x_1}\partial_{x_2}\la \theta_1\theta_2\ra=gD(r_{12})/2$, where we have abbreviated $\theta_j\equiv \theta(r_j)$ and $r_{ij}\equiv (x_i-x_j,\tau_i-\tau_j)$.
We then consider the Schwinger function
\begin{eqnarray}
&&F_{\rho_0\rho_0}\equiv\pi^{-4}\la \theta_1\theta_2\theta_3\theta_4 \ra  \\
&&~~=\frac{g^2}{4\pi^4}[D(r_{12})D(r_{34})+D(r_{13})D(r_{24})+D(r_{14})D(r_{23})]~, \nonumber
\end{eqnarray}
by Wick's theorem.
To obtain the functions after analytical continuation, we have 
\begin{equation}
D(x,\tau\rightarrow it\pm0^{+})=M(x,t)\mp iN(x,t)~,
\end{equation}where
\begin{eqnarray}
M(x,t)&=&\int_0^{\Lambda}dk k[2f_B(v_Bk)+1]\cos(kx)\cos(v_Bkt) \nonumber \\ 
N(x,t)&=&\int_0^{\Lambda}dk k\cos(kx)\sin(v_Bkt)~.
\end{eqnarray}
We also note that $D(0)=0$ and $N(x,-t)=-N(x,t)$.
(Again, one can also consider the soft-cutoff regularization, with integrand $k=0$ to $\infty$ with an extra factor $e^{-\alpha k}$.)

We therefore obtain $C_{\rho_0\rho_0}(\ell,t)$ via analytical continuation (with the order $\epsilon_1>\epsilon_2>\epsilon_3>\epsilon_4 \rightarrow 0$) as 
\begin{widetext}
\begin{eqnarray}
2C_{\rho_0\rho_0}(\ell,t)&=&F_{\rho_0\rho_0}(x_1\!=\!x_4\!=\!0,x_2\!=\!x_3\!=\!x;\tau_1\!=\!\epsilon_1,\tau_2\!=\!\epsilon_2\!+\!it,\tau_3\!=\!\epsilon_3\!+\!it,\tau_4\!=\!\epsilon_4) \nonumber \\
&+&F_{\rho_0\rho_0}(x_1\!=\!x_4\!=\!x,x_2\!=\!x_3\!=\!0;\tau_1\!=\!\epsilon_1\!+\!it,\tau_2\!=\!\epsilon_2,\tau_3\!=\!\epsilon_3,\tau_4\!=\!\epsilon_4\!+\!it) \nonumber \\
&-&F_{\rho_0\rho_0}(x_1\!=\!x_3\!=\!x,x_2\!=\!x_4\!=\!0;\tau_1\!=\!\epsilon_1\!+\!it,\tau_2\!=\!\epsilon_2,\tau_3\!=\!\epsilon_3\!+\!it,\tau_4\!=\!\epsilon_4\!) \nonumber \\
&-&F_{\rho_0\rho_0}(x_1\!=\!x_3\!=\!0,x_2\!=\!x_4\!=\!x;\tau_1\!=\!\epsilon_1,\tau_2\!=\!\epsilon_2\!+\!it,\tau_3\!=\!\epsilon_3,\tau_4\!=\!\epsilon_4\!+\!it) \nonumber \\
&=&\frac{g^2}{\pi^4}N^2(x,t)~.
\end{eqnarray}
\end{widetext}

For the soft-cutoff version, we have 
\begin{equation}
N(x,t)=\frac{\alpha(vt-x)}{[(vt-x)^2+\alpha^2]^2}+\frac{\alpha(vt+x)}{[(vt+x)^2+\alpha^2]^2}~,
\end{equation}
hence recovering the result in Ref.~\cite{doraOutofTimeOrdered2017}.
For the hard-cutoff version, we have Eq.~(\ref{eqn:Nxt}) in the main text.

Next we calculate $C_{WW}(\ell,t)$. 
The Schwinger function to consider in this case is 
\begin{eqnarray}
F_{WW}&=&\la \exp(2i(p_1\theta_1+p_2\theta_2+p_3\theta_3+p_4\theta_4)) \ra \nonumber \\
&=& \exp[2g\sum_{i<j}p_ip_jK(r_{ij})]~,
\end{eqnarray}
and the analytical continuation $K(x,\tau=it\pm 0^{+})=H(x,t)\mp iG(x,t)$.
We expand 
\begin{eqnarray}
C_{WW}(\ell,t)&=&C_{V_2V_{-2}}(\ell,t)+C_{V_{-2}V_{2}}(\ell,t) \nonumber \\
&+&C_{V_{-2}V_{-2}}(\ell,t)+C_{V_2V_2}(\ell,t)\nonumber \\
&+& e^{-i4\pi\rho_0\ell}C_{a}(\ell,t) \nonumber \\
&+& e^{i4\pi\rho_0\ell}C_{b}(\ell,t)~,
\end{eqnarray}
where
\begin{eqnarray}
C_{a}(\ell,t)&=&\frac{1}{2}\la[V_{-2}(\ell,t),V_2(0)]^{\dagger}[V_2(\ell,t),V_{-2}(0)]\ra \nonumber \\
C_{b}(\ell,t)&=&\frac{1}{2}\la[V_{2}(\ell,t),V_{-2}(0)]^{\dagger}[V_{-2}(\ell,t),V_{2}(0)]\ra \nonumber
\end{eqnarray}
Suitable combinations of the analytical continuations give us 
\begin{eqnarray}
C_{V_{-2}V_{2}}(\ell,t)&=&C_{V_{-2}V_2}(\ell,t)=C_{V_{-2}V_{-2}}(\ell,t)=C_{V_2V_2}(\ell,t)\nonumber \\
&=&2\sin^2(2gG(\ell,t))~, 
\end{eqnarray}
and
\begin{eqnarray}
C_{a}(\ell,t)&=&C_{b}(\ell,t) \nonumber \\
&=&2\exp[-8gH(\ell,t)]\sin^2[2gG(\ell,t)]~. \nonumber
\end{eqnarray}
So we have 
\begin{eqnarray}
C_{WW}(\ell,t)&=&2[4+2\cos(4\pi\rho_0\ell)\exp(-8gH)]\sin^2(2gG)~. \nonumber
\end{eqnarray}
Note that for the soft-cutoff version, 
\begin{equation}\label{eqn:Gxt_soft}
G(\ell,t)=\frac{1}{2}(\arctan(\frac{vt+\ell}{\alpha})+\arctan(\frac{vt-\ell}{\alpha}))~,
\end{equation}
recovering the result of $C_{V_{-2}V_2}(\ell,t)$ in Ref.~\cite{doraOutofTimeOrdered2017}.

Finally, we present the calculations of $C_{n_0w}(\ell,t)$, $C_{Wn_0}(\ell,t)$, $C_{\rho_0\rho_0WW}(\ell,t)$, $C_{\rho_0WW\rho_0}(\ell,t)$.
The relevant Schwinger function in this case is 
\begin{eqnarray}
F_{\rho_0\rho_0WW} &=& \pi^{-2}\partial_{x_1}\partial_{x_2}\la\theta_1\theta_2W(r_3)W(r_4) \ra \nonumber \\
&=&\frac{g}{\pi^2} D(r_{12}) \exp[2gF(r_{34})] \cos(2\pi\rho_0 x_{34}) \nonumber \\
\end{eqnarray}

We therefore obtain, upon analytical continuations,
\begin{eqnarray}
C_{\rho_0W}(\ell,t)&=&C_{W\rho_0}(\ell,t)=0 \nonumber \\
C_{\rho_0\rho_0WW}(\ell,t)&=&\frac{-4g}{\pi^2}\cos(2\pi\rho_0\ell)e^{-2gH}N(\ell,t)\sin^2(2gG(\ell,t)) \nonumber \\
C_{\rho_0WW\rho_0}(\ell,t)&=&0~. \nonumber
\end{eqnarray}
Collecting all the pieces, we obtain the result Eq.~(\ref{eqn:Cnn_luttinger}) in the main text.

\section{Comparison of $G(\ell, t)$ in hard-cutoff and soft-cutoff regularizations}\label{app:Gxt_soft_hard}
For the readers' benefit, here we collect and compare the behavior of the function $G(\ell, t)$ in different regions of interest for the two regularization schemes.
We denote $G(\ell, t; \Lambda)$ the function defined via the hard-cutoff regularization and given in Eq.~(\ref{eqn:Gxt}), while $G(\ell, t; \alpha)$ is the function defined via the soft-cutoff regularization, given in Ref.~\cite{doraOutofTimeOrdered2017} or Eq.~(\ref{eqn:Gxt_soft}).

In the short-time region $t \ll 1/(v_B \Lambda)$ or $t \ll \alpha/v_B$, we have the behavior 
\begin{eqnarray*}
G(\ell, t; \Lambda) &\sim& v_B t \frac{\sin(\Lambda \ell)}{\ell} \\
&+& (v_B t)^3 \frac{2\Lambda\ell \cos(\Lambda \ell) + (-2 +\Lambda^2\ell^2) \sin(\Lambda \ell)}{6 \ell^3} ~; \\
G(\ell, t; \alpha) &\sim& v_B t \frac{\alpha}{\alpha^2 + \ell^2} - (v_B t)^3 \frac{\alpha(\alpha^2 - 3\ell^2)}{3(\alpha^2 + \ell^2)^3} ~.
\end{eqnarray*}

In the region where one follows the rays $\ell = v t$, and $v > v_B$, or around the wavefront,
\begin{eqnarray*}
G(\ell, t; \Lambda) &\sim& \frac{\pi}{4} + \frac{1}{2} \text{Si}[\Lambda(v_B t - \ell)] ~; \\
G(\ell, t; \alpha) &\sim& \frac{\pi}{4} + \frac{1}{2} \arctan\left (\frac{v_B t - \ell}{\alpha} \right) ~,
\end{eqnarray*}
both showing non-broadening wavefront behavior.

Finally, in the long-time region, $t \gg \ell/v_B$,
\begin{eqnarray*}
G(\ell, t; \Lambda) &\sim& \frac{\pi}{2} - \frac{\cos(\Lambda v_B t) \cos(\Lambda \ell)}{\Lambda v_B t} \\
&-& \frac{\sin(\Lambda v_B t)\cos(\Lambda \ell)}{(\Lambda v_B t)^2} ~; \\
G(\ell,t;\alpha)&\sim& \frac{\pi}{2}-\frac{\alpha }{v_B t} +O(t^{-3}) ~.
\end{eqnarray*}

We therefore see that in all the cases of interest, the hard-cutoff and soft-cutoff expressions are not qualitatively different except for oscillating factors for the hard cutoff. 
We used the hard cutoff to compare with our fully controlled lattice calculation in the long-range hopping model, since such cutoff mimics the finite band width in the lattice models.

\section{Calculation of the boson-boson OTOC in the lattice models} \label{app:boson_lattice}
To compute the commutator function, it suffices to calculate the OTOC $F_{XX}(\ell, t) = \la X_0(t) X_\ell(0) X_0(t) X_\ell(0) \ra$.
For simplicity, we denote the left end-point of the lattice as $e \equiv -L/2 \!+\! 1$.
Defining the fermions $A_j = c_j^\dagger + c_j$ and $B_j = c_j^\dagger - c_j$, we can express 
\begin{eqnarray}
F_{XX}(\ell, t) = \la (A_e(t) &\dots& A_0(t) B_e(t)  \dots B_{-1}(t) \nonumber \\
A_e &\dots& A_\ell B_e \dots B_{\ell-1})^2 \ra ~,
\end{eqnarray}
and calculate it by Wick's theorem.

We need the following two-point correlation functions involving operators $A$ and $B$:
\begin{eqnarray*}
\la A_n(t) A_m \ra &=& -\la B_n(t) B_m \ra \\
&=& \frac{2}{L+1} \sum_{k} \sin(k n) \sin(k m) \\
&\times& [\cos(\epsilon_{k} t) - i \sin(\epsilon_{k} t) \tanh(\frac{\beta\epsilon_{k}}{2})] ~, \\
\la A_n(t) B_m \ra &=& -\la B_n(t) A_m \ra \\
&=& \frac{2}{L+1} \sum_{k} \sin(k n) \sin(k m) \\
&\times& [\cos(\epsilon_{k} t) \tanh(\frac{\beta\epsilon_{k}}{2}) - i \sin(\epsilon_{k}t)] ~,
\end{eqnarray*}
where the summation is running through the set ${k=p\pi/(L+1), p=1 \dots L}$.

We define $[{\tt AA}](t)^{n=a:b}_{m=c:d}$ as a matrix with matrix elements $\la A_n(t) A_m \ra$, having row index $n$ from $a$ to $b$ and column index $m$ from $c$ to $d$, and similarly for $[{\tt AB}](t)$, $[{\tt BA}](t)$, and $[{\tt BB}](t)$.
We will need also $t=0$ correlation functions $[{\tt AB}](0)$ and $[{\tt BA}](0)$, which we will denote as $[{\tt AB}]$ and $[{\tt BA}]$, i.e., by simply omitting the time argument.
We also denote the identity matrix as $[{\tt I}]$ and the zero matrix as $[{\tt 0}]$, with their sizes specified implicitly according to the context.

Now define matrices
\begin{eqnarray}
S=
\begin{pmatrix}
[{\tt 0}] & [{\tt AB}]^{n=e:0}_{m=e:-1} &[{\tt AA}](t)^{n=e:0}_{m=e:\ell} & [{\tt AB}](t)^{n=e:0}_{m=e:\ell-1} \\
- & [{\tt 0}] & [{\tt BA}](t)^{n=e:-1}_{m=1:\ell} & [{\tt BB}](t)^{n=e:-1}_{m=1:\ell-1} \\
- & - & [{\tt 0}] & [{\tt AB}]^{n=e:0}_{m=e:-1} \\
- & - & - & [{\tt 0}]
\end{pmatrix} ~, \nonumber
\end{eqnarray}
where the rest of the matrix elements are defined such that $S^T = -S$, and 
\begin{widetext}
\begin{eqnarray}
R=
\begin{pmatrix}
[{\tt I}] & [{\tt AB}]^{n=e:0}_{m=e:-1} &[{\tt AA}](t)^{n=e:0}_{m=e:\ell} & [{\tt AB}](t)^{n=e:0}_{m=e:\ell-1} \\
[{\tt BA}]^{n=e:-1}_{m=e:0} & -[{\tt I}] & [{\tt BA}](t)^{n=e:-1}_{m=1:\ell} & [{\tt BB}](t)^{n=e:-1}_{m=1:\ell-1} \\
[{\tt AA}](-t)^{n=e:\ell}_{m=e:0} & [{\tt AB}](-t)^{n=e:\ell}_{m=e:-1} & [{\tt I}] & [{\tt AB}]^{n=e:0}_{m=e:-1} \\
[{\tt BA}](-t)^{n=e:\ell-1}_{m=e:0} & [{\tt BB}](-t)^{n=e:\ell-1}_{m=e:-1} & [{\tt BA}]^{n=e:\ell-1}_{m=1:\ell} & -[{\tt I}]
\end{pmatrix}~. \nonumber
\end{eqnarray}
\end{widetext}
Using Wick's theorem, we then have 
\begin{equation}
F_{XX} = \text{Pf}
\begin{bmatrix}
S & R \\
-R^T & S
\end{bmatrix} ~,
\end{equation}
where $\text{Pf}[Q]$ evaluates the Pfaffian of an antisymmetric matrix $Q$.

%

\end{document}